%% file: subgrid_scheme.tex
\documentclass[authoryear,1p]{elsarticle} %
\usepackage{times}

\input{usepackages}
\input{newcommands}

\begin{document}

\title{A Dual Grid Method for the Compressible Two-Fluid Model which
Combines Robust Flux Splitting Methodology 
with High-Resolution Capturing of Incompressible Dynamics}

\author{A.H.~Akselsen\corref{cor1}}
\address{Department of Energy and Process Engineering, Norwegian University of Science and Technology, Kolbj{\o}rn Hejes v.\ 1B, 7491 Trondheim, Norway}
\ead{andreas.h.akselsen@ntnu.no}

\cortext[cor1]{Corresponding author}

\begin{keyword}
dual grid method\sep hybrid central-upwind scheme\sep two-phase flow\sep pipe flow\sep two-fluid model\sep Roe scheme
\end{keyword}%


 
\begin{abstract}
\input{abstract}

\end{abstract}
\maketitle

\section{Introduction}
\label{sec:introduction}
Dynamic flow simulators have been a vital tools in industries like the nuclear industry~\citep{RELAP5_manual,Barre_CHATHARE} and the petroleum industry~\citep{Bendiksen_OLGA,Larsen_PeTra}.
Predictions of the flow topology, fluid transport and pressure loss, as well as the simulation of potentially damaging or dangerous scenarios, are among the key features of these simulation tools.
Many, if not most, such simulators are based on the so-called two-fluid model, which is derived by averaging the fundamental conservation equations over district flow fields -- for example a gas and a liquid field.
The mechanism for hydrodynamic growth of long-wavelength instabilities is known to be an inherent feature of the two-fluid model~\citep{Barnea_VKH_and_IKH}.
Capturing this mechanism has in recent years become a popular method for predicting the flow field topology and flow regime transitions in dynamic simulators~\citep{Issa_two_phase_capturing}.
This methodology has even found its way into commercial software for large scale pipeline systems~\citep{Kjolaas_Leda_capturing,RELAP5_manual}.
\\

The characteristic speeds active in gas-liquid flows can differ by several orders of magnitude. 
Sonic waves are artefacts of fluid compressibility and propagate much quicker than hydraulic waves pertaining to changes in the volume fraction.
Hydraulic waves are responsible for the hydrodynamic growth of long-amplitude surface waves, often of primary interest.
Acoustic or `sonic' wave (waves in pressure) work on a time scale too small to affect long-amplitude waves significantly, giving the pressure an idle role regarding surface waves. 
All the same, sonic waves must be computed carefully if the simulation procedure is to remain numerically stable, placing a strict time step restriction on explicit solvers.
\\

A simple way to allow a numerical scheme to operate at the slow time scales suited for hydraulic waves is to ignore compressibility altogether. 
The four-equation two-fluid model can then be reduced to a two-equation form as two pressure waves are removed from the system.
Worth mentioning in regard to this incompressible two-equation model is
\cite{Keyfitz}, who analysed it mathematically.
\cite{Wangensteen} proposed a flux splitting technique for intermittent single phase -- two phase flows (slug flow) built on the two-equation model.
He also constructed a Roe scheme based on Keyfitz' formulation.
The incompressible two-fluid model has further been used by \cite{Holmaas_roll_wave_model} to effectively simulate high-pressure flow in the roll-wave regime. 
These simulations compared favourably to the experimental campaign of \cite{George_PhD}.
Holm{\aa}s' formulation of the incompressible  model was based on the formulation used by \cite{Watson_wavy}, 
which is somewhat cleaner than the one investigated by Keyfitz.
The present author used that model formulation to construct a Roe scheme and schemes based on the principle of characteristics~\citep{Akselsen_char_Roe}, proving very efficient. 
\\

The present work revisits the dual grid methodology for resolving hydraulic and acoustic waves on separate grids.
The concept was investigated in~\citep{Akselsen_dual_grid} using a primitive decoupling that discriminated between gas and liquid phases.
The gas phase was associated with acoustic waves and designated to a coarse grid, while the liquid phase was resolved in greater detail.
Although the method indicated the potential of the dual grid strategy, it suffered from grid dependent disturbances 
generated as large scale hydraulic information was projected down onto the smaller scales.

The presently presented method distinguishes between scale based on the compressibility itself, 
projecting only information pertaining to compressibility down onto the smaller scales.
A two-way coupling between the two computational grids is achieved using the flux splitting approach due to \cite{Evje_HCU_schemes}, 
which also ensures a robust yet explicit treatment of the pressure. 
At the same time, the presented dual grid scheme exploits the benefits of the incompressible two-fluid model, in particular its neatness and simple eigenstructure. 
The resulting method is one which successfully neutralizes the difference in sonic and hydraulic travelling speeds.
It allows for simple, explicit and affordable simulation in a variety of cases which on a single grid arrangement would require a semi-implicit formulation or significant computer power. 
\\

This article is structured as follows:
The four-equation two-fluid model for stratified pipe flow is briefly presented in \autoref{sec:two_fluid_model}.
\autoref{sec:dual_grid_scheme} provides the building blocks of the dual grid scheme.
This includes a summary of the Hybrid Central-Upwind (\HCU{}) flux splitting scheme (\autoref{sec:HCU},) 
the incompressible two-fluid model with a Roe scheme discretization (\autoref{sec:Roe},)
and the means
by which these are coupled
(Subsections~\ref{sec:dual_grid_scheme:integral_to_sub} and \ref{sec:dual_grid_scheme:sub_to_integral}.)
Linear stability expressions are presented in \autoref{sec:vonNeumann}, while
\autoref{sec:contact_discont} 
illustrates how the two-grid arrangement maintains the properties of the \HCU{} scheme presented in \citep{Evje_HCU_schemes}.
\added{
Numerical tests presented in \autoref{sec:exp} are given in two parts. 
Two basic benchmark tests from original \HCU{} publication (shock tube and water faucet) are repeated with extra subgrid resolution in \autoref{sec:exp:dispersive_tests}.
Two larger problems, more closely related to engineering, are studied in \autoref{sec:exp:stratified_tests}. 
Acoustic-hydraulic wave interactions, computational efficiency and flow regime prediction are considered in these problems. 
}
A summary is given in \autoref{sec:conclusions}.

\section{The two-fluid model for stratified pipe flows}
\label{sec:two_fluid_model}

\begin{figure}[h!ptb]%
\centering
\begin{minipage}[c]{.3\columnwidth}
\includegraphics[width=\columnwidth]{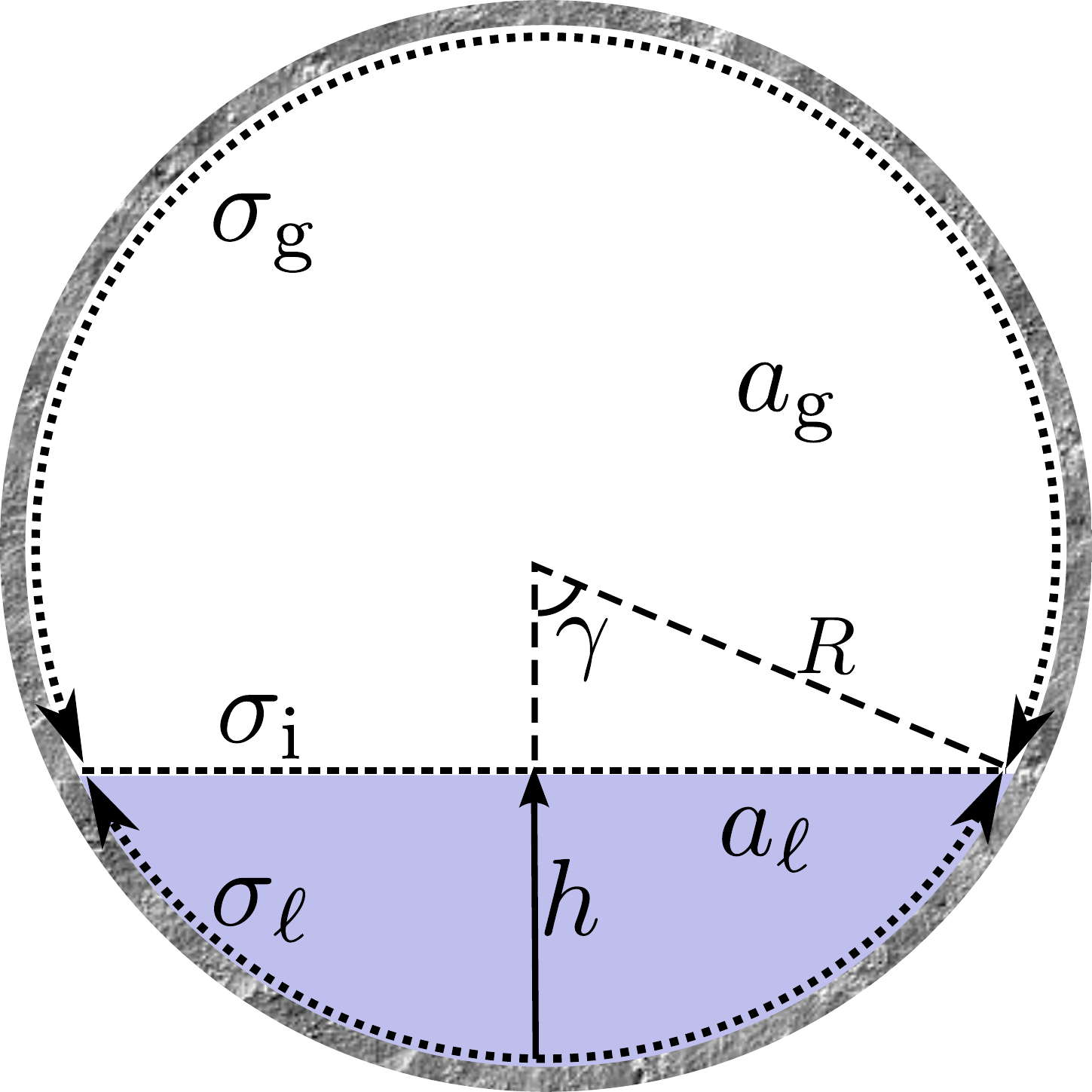}%
\end{minipage}
\hfill
\begin{minipage}[c]{.55\columnwidth}
\includegraphics[width=\columnwidth]{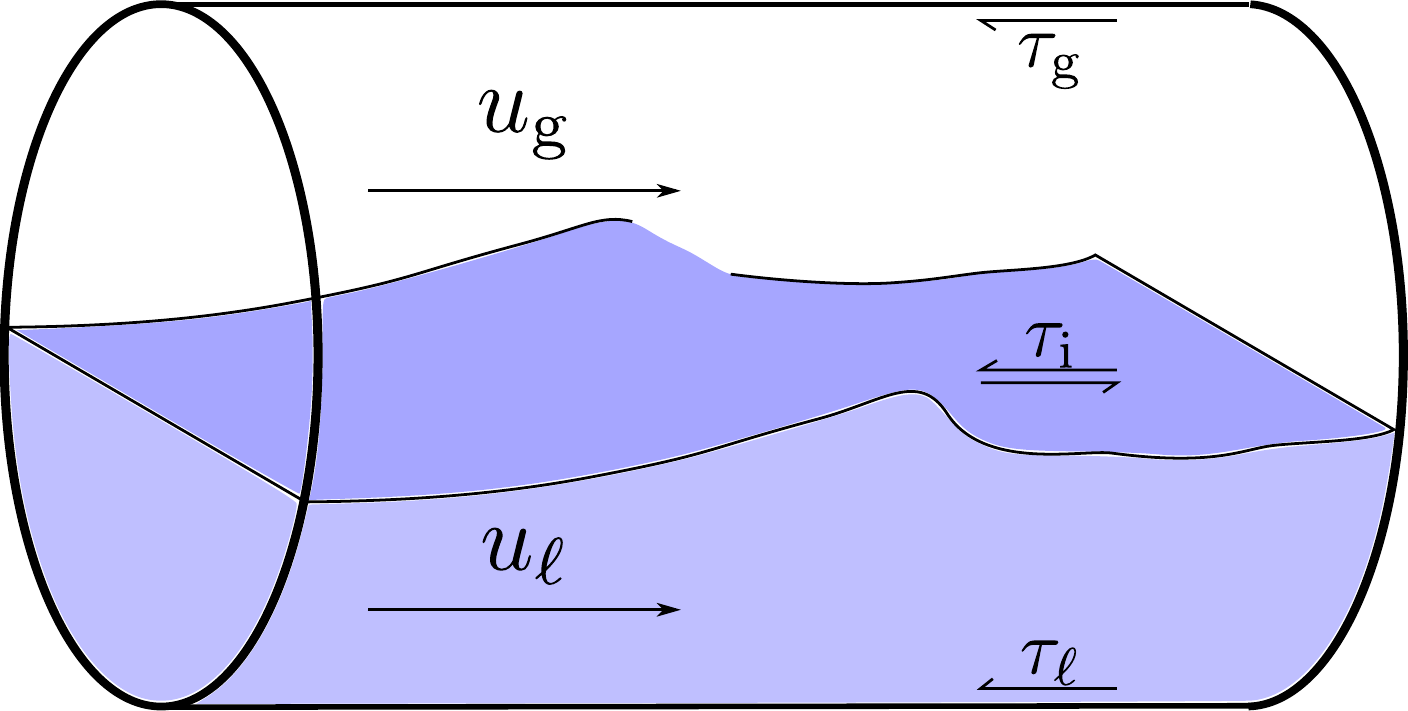}%
\end{minipage}
\caption{Pipe cross-section}%
\label{fig:cross_section}%
\end{figure}

The compressible, equal pressure four-equation two-fluid model for stratified pipe flow results from an averaging of the conservation equations within a flow field over the pipe cross-section.  
It is commonly written
\begin{subequations}
\begin{align}
	&\pp_t m\k + \pp_x i\k = 0,  \label{eq:base_model_with_pressure:mass} \\
	&\pp_t i\k + \pp_x ( u\k i\k ) + a\k \pp_x p + g_y m\k  \pp_x h = s\k,  \label{eq:base_model_with_pressure:mom}\\
	&\al+\ag=\area, \label{eq:sum_a}\\
	& \rho\k = \rho\k\of{p}. \label{eq:EOS}
\end{align}%
\label{eq:base_model_with_pressure}%
\end{subequations}%
\deleted{Specific} Mass $m\k = \rho\k \ak$ and momentum $i\k=\rho\k \ak u\k$ \added{per unit length} are conserved properties.
Field $\phaseindex$, occupied by either gas, $\phaseindex = \mr g$, or liquid, $\phaseindex = \ell$, is segregated from the other field. 
$p$ is here the pressure at the interface, assumed the same for each phase as surface tension is neglected.
$h$ is the height of the interface from the pipe floor, and the term in which it appears originates from approximating a hydrostatic wall-normal pressure distribution. 
$u\k$ and $\rho\k$ are the mean fluid velocity and density, respectively, in field $\phaseindex$. 
The momentum sources are 
$s\k = -\tau\k \sigma\k \pm \tau\_i \sigma\_i - m\k g_x$,
where $\tau\k$ and $\sigma\k$ is the skin friction and perimeter of the pipe wall in field $k$, respectively.
$\tau\_i$ and  $\sigma\_i$ refer to the interphase; see \autoref{fig:cross_section}.
$g_x = g\sin\theta$ and $g_y = g\cos\theta$ are the horizontal and vertical components of the gravitational acceleration, respectively.
$\theta$ is here the pipe inclination, positive above datum.
\\

The circular pipe geometry governs the relationship between the level height $h$, the \replaced{field }{specific} areas $\ak$ and the perimeter lengths $\sigma\k$ and $\sigma\_i$.
These are algebraically interchangeable through a geometric function
\begin{equation}
h = \HofAl\of \al
\label{eq:Alofh}
\end{equation}
whose derivative is  $\dHdAl = 1/\sigma\_i$.
See \egg\ \citep{Akselsen_char_Roe} for expressions of the geometric relationships.
\added{$\mc A$ is the total cross-section area.}

Linear equations of state
\begin{equation}
\rho\k\of{p} = \rho_{k,0} + \drhokdpEOS\br{p - p\_{0}}
\label{eq:EOS_linear}
\end{equation}
are used in this work.
Excluding energy equations and any temperature dependency in the equations of state makes this flow model isentropic.

Finally, a \textit{pressure evolution equation} may be obtained from the model \eqref{eq:base_model_with_pressure} by summing both mass equations and applying the chain rule to the time differential. Using \eqref{eq:sum_a} and \eqref{eq:EOS}, we obtain
\begin{equation}
\pp_t p + \br{\rho\g \pp_x i\l + \rho\l \pp_x i\g}\Fkappa = 0,
\label{eq:pressure_evolution}
\end{equation}
where
\begin{equation}
\Fkappa = \frac1{\rho\g \al \drholdp + \rho\l \ag \drhogdp}.
\label{eq:Fkappa}
\end{equation}
The mixture speed of sound may be approximated by
$
\sqrt{\br{\rho\l \ag+\rho\g \al}\kappa}
$
\citep{Flaatten_common_two_fluid}.

\section{The dual grid scheme}
\label{sec:dual_grid_scheme}
The dual grid scheme solves the same set of transport equations on two separate grids,
the purpose of which is to allow acoustic waves (waves in pressure) and hydraulic waves (waves in volume fraction) to be treated numerically at differing length scales.  
The grid on which the compressible two-fluid model is solved will be term the \textit{principal grid}.
Alongside this, an incompressible two-fluid model is solved on a finer grid which we shall term the \textit{subgrid}.
The purpose of the subgrid computations is to capture the hydraulic surface flow details. 
The purpose of the principal grid computations is to account for compressibility, pressure waves and to maintain numerical conservation.
Information contained in the two grids  needs to be coupled for the evolution of the principal and subgrid models to remain consistent.

\begin{figure}[h!ptb]%
\centering
\includegraphics[width=.85\columnwidth]{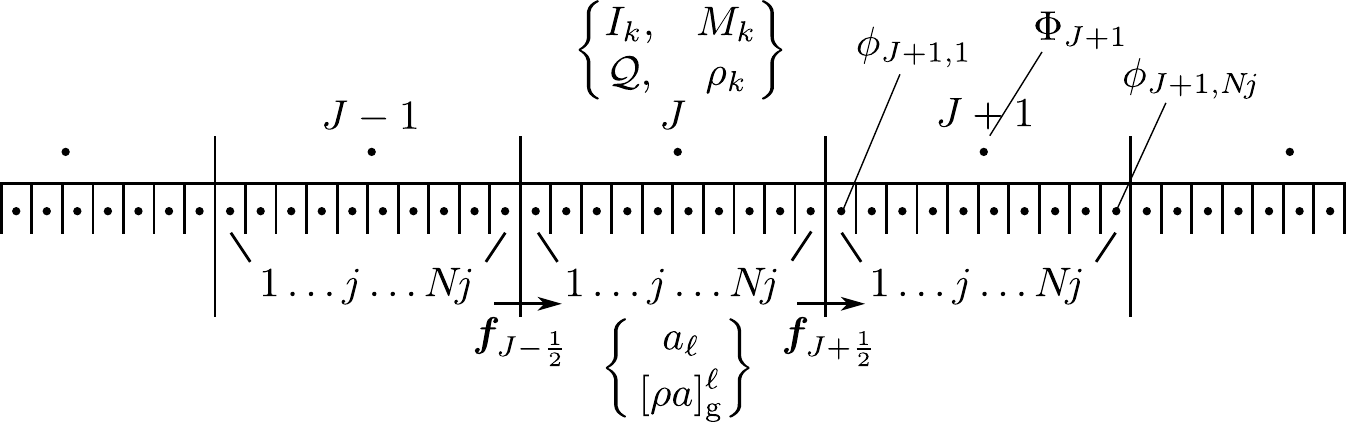}%
\caption{Grid arrangement of the dual scale scheme}%
\label{fig:dual_grid_scheme}%
\end{figure}

\autoref{fig:dual_grid_scheme} shows a schematic of the two grids in the dual grid scheme. 
Upper-case symbols will in the following be reserved for discrete variables associated with the principal grid, lower-case symbols indicating subgrid variables. 
Likewise, upper-case indexation $J$ generally refer to cells of the principal grid, while lower-case indices $j$ pertain to the subgrid cells. 
Double indexation $\phi_{J,j}$ will be used whenever we wish to specify the position of a subgrid cell in relation to the principal grid, but not consistently. 
The subgrid domain is divided with an integral number of subgrid cells $\Nj{}$ per principal grid cell. For simplicity, we make both grids uniform with $\Nj{}$ constant. 
The total number of principal grid cells is denoted $\NJ{}$, and the total number of subgrid cells is then $\NJ{}\!\times\!\Nj{}$.
\\

In the following, the numerical schemes of the principal grid and the subgrid are presented, followed by a description of the method employed to couple information from the two.

\subsection{Principal Grid -- The Hybrid Central-Upwind Scheme}
\label{sec:HCU}%
A finite volume differentiation of \eqref{eq:base_model_with_pressure} is written in terms of conserved variables
\begin{subequations}
\begin{align}
\frac{M\kJ\nn-M\kJ\n}{\dt} &+ \frac{\F\kJph\n-\F\kJmh\n}{\dX} = 0,
\label{eq:base_model_with_pressure_discretized:mass}%
\\
\frac{I\kJ\nn-I\kJ\n}{\dt}  
&+ \frac{(U I)\kJph\n-(U I)\kJmh\n}{\dX}\nonumber \\
&+ \A\kJ \frac{P\Jph\nn-P\Jmh\nn}{\dX}  \nonumber  \\
&+ g_y M\kJ\n \frac{H\Jph\n-H\Jmh\n}{\dX} = S\kJ\n. \label{eq:base_model_with_pressure_discretized:mom}
\end{align}%
\label{eq:base_model_with_pressure_discretized}%
\end{subequations}%
The Hybrid Central-Upwind (HCU) scheme is used for the terms in \eqref{eq:base_model_with_pressure_discretized}, now briefly presented.
See \citep{Evje_HCU_schemes} for more details. 

\subsubsection*{Pressure Differentiation}
Acoustic terms are treated with Lax-Friedrich discretization, which is centred and numerically viscous.
Thus formulating the pressure evolution equation \eqref{eq:pressure_evolution} in a control volume placed over the cell face $J+\h$ yields
\begin{align}
\label{eq:Pnn}
P\Jph\nn = &\frac12\br{P\Jp\n +P\J\n} - \frac{\dt}{\dx} \Fkappa\Jph\n \\
& \times\sqbrac{\rho_{\mr g,J+\h}\n \br{I_{\ell,J+1}\n-I_{\ell,J}\n} + \rho_{\ell,J+\h}\n\br{I_{\mr g,J+1}\n-I_{\mr g,J}\n}}. \nonumber
\end{align}

\subsubsection*{Flux Splitting}
Flux splitting in the \HCU{} scheme is based on the way in which the phase fraction is a function of the local pressure $p$ and \replaced{linear mass densities $m\k$ }{specific mass}. Precisely, differentiating $m\k=\ak\rho\k$ gives
\begin{equation}
\dd m\k = \ak \drhokdp \dd p + \rho\k \dd \ak.
\label{eq:dm}
\end{equation}
Inspired by \eqref{eq:dm}, 
convective fluxes $\F\k=\rho\k \A\k U\k$ are split into a sonic flux $\Fp$, related to pressure changes, and a hydraulic flux $\Fal$, related to changes in volume fraction, as follows: 
\begin{subequations}
\begin{align}
\F\l = \Al \drholdp \Fp + \rho\l \Fal, \label{eq:Fl} \\
\F\g = \Ag \drhogdp \Fp - \rho\g \Fal. \label{eq:Fg}
\end{align} %
\label{eq:Fk_splitting}%
\end{subequations}
We wish to differentiate the convective fluxes $\F\k$ in such a way that $\Fp$ receives sufficient numerical diffusion for robustness while retaining a high accuracy in $\Fal$.
Rearranging \eqref{eq:Fk_splitting} in terms of $\Fp$ and $\Fal$, we write
\begin{subequations}
\begin{align}
\Fp &= \br{\rho\g \F\l\CC + \rho\l \F\g\CC} \label{eq:F_p} \Fkappa, \\
\Fal &= \br{\drhogdp \Ag \F\l\UU + \drholdp \Al \F\g\UU} \Fkappa. \label{eq:F_al}
\end{align}%
\label{eq:flux_splitting}%
\end{subequations}%
The `C' appended to the phase fluxes in \eqref{eq:F_p} indicates that these are to be computed in a centred manner, with numerical viscosity sufficient to ensure stability in the sonic features of the mass transport.
Likewise, the `U' added to the phase fluxes in \eqref{eq:F_al} indicates that an upwind discretization is to be applied, providing 
higher accuracy in the flux component attributed to slower hydraulic waves.
Later, we will see how this flux splitting can be utilized further in the dual grid scheme by associating the sonic fluxes with a spatial scale which is long in comparison to the spatial scale of the hydraulic fluxes. 

Inserting \eqref{eq:flux_splitting} back into \eqref{eq:Fk_splitting} now yields
\begin{subequations}
\begin{align}
\F\l &= \br{  \rho\g \Al \drholdp \F\l\CC + \rho\l \Ag \drhogdp  \F\l\UU + \rho\l \Al \drholdp\br{ \F\g\CC - \F\g\UU } }\Fkappa, \label{eq:Fl_disc} \\
\F\g &= \br{  \rho\l \Ag \drhogdp \F\g\CC + \rho\g \Al \drholdp  \F\g\UU + \rho\g \Ag \drhogdp\br{ \F\l\CC - \F\l\UU } }\Fkappa.  \label{eq:Fg_disc}
\end{align}%
\label{eq:flux_splitting_result}%
\end{subequations}%
Cell face subscripts $J+\frac12$ are here implied as \eqref{eq:flux_splitting_result} is formulated at the cell faces of the principal grid.
Note that \eqref{eq:flux_splitting_result} is consistent, \ie, it reduces to simple equalities as $\F\k\CC,\F\k\UU\rightarrow \F\k$.
\\

Central difference convective fluxes $\F\k\CC$ have been related to the pressure evolution and should therefore be differentiated in a consistent way. 
The pressure evolution equation \eqref{eq:Pnn} is expressed with Lax-Friedrich differentiation in a control volume 
covering the cell face $J+\h$.
Convective fluxes of the pressure control volume are thus located at cell centres.
Expressing cell face fluxes $\F\kJph\CC$ as the arithmetic mean of these, the modified  Lax-Friedrich fluxes
\begin{equation}
\F\kJph\CC = \frac12 \br{\F\kJp+\F\kJ} - \frac14 \frac\dx\dt \br{M\kJp-M\kJ}
\label{eq:F_C}
\end{equation} 
are obtained.
Upwind fluxes were in \citep{Evje_HCU_schemes} chosen
\begin{equation}
\F\kJph\UU = 
\begin{cases}
M\kJ \ol{U\kJph} & \text{if}\quad \ol{U\kJph} > 0,\\
M\kJp \ol{U\kJph} & \text{otherwise}
\end{cases}
\label{eq:F_U}
\end{equation}
with $\ol{U\kJph} = \h\br{U\kJp+U\kJ}$.
Momentum convection is in the \HCU{} scheme the natural extension of \eqref{eq:F_U}, namely
\begin{equation}
(UI)\kJph = 
\begin{cases}
I\kJ \ol{U\kJph} & \text{if}\quad \ol{U\kJph} > 0,\\
I\kJp \ol{U\kJph} & \text{otherwise}.
\end{cases}
\label{eq:UI_Jph_HCU}
\end{equation}

\subsection{The Incompressible Two-Fluid Model}
\label{sec:incompressible_model}
We will use an incompressible two-equation formulation of the two-fluid model \eqref{eq:base_model_with_pressure} as the hydraulic model for the subgrid.
The incompressibility of this model eliminates acoustic waves so that fine subgrids can be used without needing reduced time steps.

Reducing the momentum equations of \eqref{eq:base_model_with_pressure} with their respective mass equations and eliminating the pressure term between them results in
\begin{equation}
\partial_t \bv + \partial_x \bf = \bs + \bm e,
\label{eq:base_model}
\end{equation}
with variables, fluxes and sources
\begin{align}
\bv &= \begin{pmatrix}
	\al\\ \diffk{\rho u}
\end{pmatrix},   
\quad
\bf = \begin{pmatrix}
	\al u\l\\ \frac12 \!\diffk{\rho  u^2 }+  \my h
\end{pmatrix}, \nonumber
\\
\bs  &= \begin{pmatrix}
	0\\ -\mx - \diffk{\frac{\tau \sigma}\a } + \tau\_i \sigma\_i \br{\frac1\al+\frac1\ag}
\end{pmatrix}. \nonumber
\intertext{The compressibility terms read}
\bm e &= \begin{pmatrix}
	\frac{\al}{\rho\l}\br{\pp_t  + u\l \pp_x}\rho\l
	\\
	\diffk{u\, \pp_t \rho+ \br{{u^2/2} + \my  h } \pp_x\rho}
\end{pmatrix}. \label{eq:e}
\end{align}
The short-hand 
\[
\diffk\cdot = (\cdot)\l-(\cdot)\g
\] 
has here been introduced.
Assuming incompressible phases results in $\bm e \equiv \bm 0$ and 
the identities
\begin{align}
\al+\ag &= \area\of x, & \ql+\qg &= \Qm\of t
\label{eq:sum_a_au}
\end{align}
then close the subgrid model. 
The latter identity in \eqref{eq:sum_a_au} has been obtained from summing the gas and liquid mass equations and applying the former identity.
Both $\area$ and $\Qm$ are parametric.

\subsection{Subgrid -- A Roe Scheme}
\label{sec:Roe}
The Roe scheme presented in \citep{Akselsen_char_Roe} will be applied to the incompressible model \eqref{eq:base_model} in the subgrid. 
In contrast to the compressible two-fluid model~\eqref{eq:base_model_with_pressure}, the incompressible model \eqref{eq:base_model} has an explicit and simple eigenstructure. 
The model also has a conservative form so that the 
linearized Riemann solver does not have to address the issue of 
choosing a state integration path, which would be necessary with a Riemann solver for the compressible model \citep{Toumi_Riemann_solver_two_fluid_model}.

Quickly summarized, the explicit Roe scheme for \eqref{eq:base_model} computes
\begin{equation}
\bv_j\nn = \bv_j\n - \frac\dt\dx \br{\bf\jph\n-\bf\jmh\n} + \dt\,\bs_j\n
\label{eq:Roe_scheme}
\end{equation}
with the fluxes 
\begin{equation}
\bf\jph = \tfrac12\br{\bf\jp+\bf_j} -\tfrac12|\Roe|\jph (\bv\jp-\bv_j).
\label{eq:Roe_sol_f}
\end{equation}
$\Roe\jph$ is here the Roe average matrix of the Jacobian $\jac$ at the cell face and
\begin{equation}
|\Roe| = 
\begin{pmatrix}
	|\lambda^+|+|\lambda^-| 			& \br{|\lambda^+|-|\lambda^-|}/\symkappa \\
	(|\lambda^+|-|\lambda^-|)\symkappa  &|\lambda^+|+|\lambda^-|  
\end{pmatrix}.
\label{eq:Lambda_LL}
\end{equation}
Eigenvalues of \eqref{eq:base_model} are
\begin{equation}
\lambda^\pm = \frac{\frac{\rho\l u\l}{\al}+\frac{\rho\g u\g}{\ag}  \pm \symkappa}{ \frac{\rho\l}{\al}+\frac{\rho\g}{\ag}},
\label{eq:eigenvalues}
\end{equation}
with
\begin{equation}
	\symkappa = \sqrt{\my \br{ \frac{\rho\l}{\al}+\frac{\rho\g}{\ag}}\dHdAl - \frac{\rho\l \rho\g}{\al \ag} \br{u\g-u\l}^2}.
	\label{eq:symkappa}
\end{equation} %
\eqref{eq:Lambda_LL} is computed with the following averages:
\begin{subequations}
\begin{align}
\a\kjph &= \tfrac12\br{\a\kjp+\a\kj},
\\
u\kjph &= \tfrac12\br{u\kjp+u\kj},
\\
\HofAl'\jph &= 
\begin{cases}
\frac{h\jp-h_j}{\aljp-\alj} &\text{if}~ \aljp\neq\alj, \\
\dHdAl\of{\a\ljph} & \text{otherwise}.
\end{cases}
\end{align}
\label{eq:Roe_mean}%
\end{subequations}%
Particularly, this Roe scheme reduces to the very simple upwind scheme
\begin{equation}
	\bf\jph = \begin{cases}
	\bf_j & \text{if}\quad \lambda^+,\lambda^- > 0,\\
	\bf\jp & \text{if}\quad \lambda^+,\lambda^- <0,\\
	\end{cases}
	\label{eq:Roe_supercritical}
\end{equation}
if both characteristic velocities are of the same sign.
See \citep{Akselsen_char_Roe} for more details on this Roe scheme.

\added{
The time step restriction of this scheme is determined by the spectral radius:}
\begin{equation}
\dt \leq \CFL{}\frac{\dx}{\max\{|\lambda_j^\pm|\}}.
\label{eq:Roe_CFL}
\end{equation}

\deleted{
Time steps are dynamically computed according the spectral radius:
}

Simpler schemes, such as a first order upwind discretization
\begin{align}
f_{1,j+\h} &=\alj \max(u\lj,0)+\aljp \min(u\ljp,0), \nonumber\\
f_{2,j+\h} &= \begin{aligned}[t]
\diffk{\a_j u_j\max(u_j,0)+\a\jp u\jp \min(u\jp,0)} \\
+\my\tfrac12(h\jp+h_j),
\end{aligned}
\label{eq:f_simple_upwind}
\end{align}
have also been found to work well in the subgrid. 
These results are however omitted for the sake of briefness.

\subsection{Coupling the grids}
\label{sec:dual_grid_scheme:couplings}
The compressibility error \eqref{eq:e} of the subgrid model \eqref{eq:base_model}, along with its primitive variable form, means that average properties of the principal grid and the subgrid may diverge with time if not appropriately coupled. 
Directly adjusting state variables in the subgrid violates the information flow of the system and usually results in numerical errors transcending from the principal grid down onto the subgrid \citep{Akselsen_dual_grid}. 
Rather than trying to force exact consistency between the grids, we settle for term-by-term couplings in the scheme equation systems which ensures either consistency or close proximity. 
The term-by-term couplings mean that the subgrid methodology can be regarded by way of a scheme extension, which does not necessitate significant alterations to a single grid implementation.

Good proximity is achieved through the following measures: 
\begin{enumerate}
	\item The model \eqref{eq:base_model_with_pressure} is restricted to compressibility in the gas phase only. 
\label{en:incomp_liq}
	\item Upwind mass fluxes $\F\k\UU$ of the principal grid is made to correspond precisely to the volumetric flux $f_1=\al u\l$ of the subgrid  where the cell faces of the two grids overlap.
\label{en:F_U_eq_f}
	\item Subgrid information is applied to the hydraulic terms of the compressible model.  
\label{en:non_conservative_terms}
\end{enumerate}
\vspace{-3mm}
By \labelref[Measure]{en:incomp_liq} the phase fraction error in \eqref{eq:e} vanishes.
It further reduces the \HCU{} liquid convection \eqref{eq:Fl_disc} to $\F\lJph = \F\lJph\UU$.
\labelref[Measure]{en:F_U_eq_f} will in turn ensure that that 
the volumetric flow entering and leaving a cell of the principal grid exactly equals the net flux through the overlapping subgrid cells. 
Analogous to the divergence theorem, we then achieve 
$\A\lJ\n = \frac1\Nj \sum_{j=1}^\Nj \a_{\ell,J,j}\n$; 
that the mean volume fractions remains equal in both grids within the domain of a principal grid cell. 

Momentum consistency between grids is not directly imposed
though close proximity of the momentum in the two grids is maintained due to the \ref{en:F_U_eq_f}nd and \ref{en:non_conservative_terms}rd measure.
These will be described in more detail in \autoref{sec:dual_grid_scheme:sub_to_integral}.
\\

\labelref[Measure]{en:incomp_liq} -- modelling the liquid phase as being incompressible -- can be a restrictive assumption.
Volume fraction inconsistencies are however believed to be small in most cases if the
liquid compressibility is weak.

\subsubsection{Coupling the Principal Model to the Subgrid Model}
\label{sec:dual_grid_scheme:integral_to_sub}
After solving the subgrid equations \eqref{eq:Roe_scheme} for $\bv=(\al,\diffk{\rho u})^T$,
new primitive variables are recovered through \eqref{eq:sum_a_au} yielding
\begin{align}
&\begin{aligned}
\al &= v_1,
\\
\ag &= \area - \al,
\end{aligned}
&
u\l &= \frac{ \rho\g \Qm +  \ag v_2}{\ag\rho\l+\al\rho\g},
&
u\g &= \frac{ \rho\l \Qm -  \al v_2}{\ag\rho\l+\al\rho\g}.
\label{eq:prim_var_comp}
\end{align}
%
Densities and the mixture flow rate, which are spatially uniform in an incompressible flow situation, 
are interpolated from the principal grid variables;
if $\tilde \rho\k\n\of x$ are interpolations of $\{\rho\kJ\n\}$ and $\tilde \Qm\n\of x$ is an interpolation of $\Big\{\frac{I\lJ\n}{\rho\lJ\n}+\frac{I\gJ\n}{\rho\gJ\n}\Big\}$, 
then 
\begin{align}%
\rho\kj\n &= \tilde \rho \k\n\of{x_j},
&
\Qm_j\n &= \tilde \Qm\n\of{x_j}.
\label{eq:interpolation}%
\end{align}%
We thus solve the `locally incompressible' subgrid model \eqref{eq:Roe_scheme}, but impose a compressible evolution indirectly by using densities and a mixture flow rate that in \eqref{eq:prim_var_comp} varies spatially.
Quadratic interpolation is applied to the tests presented herein.

Generally, interpolation onto a differential mode `from above' causes numerical problems around discontinuities. 
Here, however, we interpolate on variables which are constant in the subgrid model,
and which are computed on a coarser and more diffusive grid.
Therefore, unless a discontinuity is imposed as an initial condition,
we expect compressibility variables to appear smoothly distributed viewed from the subgrid, and thus liable to interpolation.

\subsubsection{Coupling the Subgrid Model to the Principal Model}
\label{sec:dual_grid_scheme:sub_to_integral}
We now seek to obtain a two-way coupled grid arrangement,
proposing term-by-term couplings to the compressible model \eqref{eq:base_model_with_pressure_discretized}.
\\

The subgrid liquid flow rates $f_1 = \al u\l$ from \eqref{eq:Roe_sol_f} can be used for computing upwind mass convection $\F\k\UU$ in \eqref{eq:flux_splitting_result}. 
Instead of \eqref{eq:F_U}, the fluxes
\begin{subequations}
\begin{align}
\F_{\ell,J+\h}\UU &= 
\rho\lJ \max(f_{1,J+\h},0)   \nonumber\\&\quad + \rho\lJp \min(f_{1,J+\h},0),
\label{eq:F_U_subgrid:liq}
\\
\F_{\mr g,J+\h}\UU & = 
\rho\gJ \max(\Qm_J - f_{1,J+\h},0)  \nonumber\\&\quad + \rho\gJp \min(\Qm\Jp - f_{1,J+\h},0)
\label{eq:F_U_subgrid:gas}%
\end{align}
\label{eq:F_U_subgrid}%
\end{subequations}%
are proposed.
Here, $f_{1,J+\h}$ is the volume flux at the subgrid cell face overlapping the right cell face of principal grid cell $J$, \ie,
$f_{1,J+\h}= f_{1,J,\Nj+\h} = f_{1,J+1,\h}$ according to the index notation of \autoref{fig:dual_grid_scheme}.
The flux splitting of the \HCU{} scheme is perfect for this purpose as the hydraulic component of the mass fluxes are now compounded from the appropriate scales, using a reliable Roe upwind scheme based on volume fraction characteristics. 
The sonic component of the mass fluxes is in turn confined to a longer spatial scale and is supplied with sufficient stabilization. 
As noted earlier, expression \eqref{eq:F_U_subgrid:liq} provides perfect volume fraction consistency between the principal grid and the subgrid if $\rho\l$ is constant.

Analogous to the HCU scheme, a natural extension of \eqref{eq:F_U_subgrid},
\begin{subequations}
\begin{align}
\br{UI}\lJph &= 
\rho\lJ u_{\ell,J,\Nj} \max(f_{1,J+\h},0)  \nonumber
\\& \quad+ \rho\lJp u_{\ell,J+1,1} \min(f_{1,J+\h},0),
\label{eq:IU_subgrid:liq}
\\
\br{UI}\gJph & = \rho\gJ u_{\mr g,J,\Nj} \max(\Qm_J - f_{1,J+\h},0)  \nonumber
\\& \quad+ \rho\gJp u_{\mr g,J+1,1} \min(\Qm\Jp - f_{1,J+\h},0),
\label{eq:IU_subgrid:gas}%
\end{align}
\label{eq:UI_subgrid}%
\end{subequations}%
is used for the momentum convection terms in place of \eqref{eq:UI_Jph_HCU}.
%
\\

Also the non-conservative terms and the source terms of the compressible model \eqref{eq:base_model_with_pressure_discretized} can be computed with subgrid information.
Taking the average of the skin frictions already computed in the subgrid,
\begin{align}
S\k\n = \frac1{\Nj}\sum_{j = 1}^\Nj \br{ -\tau\kj\n \sigma\kj\n \pm \tau_{\mr i,j}\n \sigma_{\mr i,j}\n - m\kj\n g_{x,j}},
\label{eq:S_from_sub}
\end{align}
is suggested. 
Finally, non-conservative level height terms may be computed
\begin{equation}
 \br{g_y M\k\pp_x H}_J  = \frac1{\Nj}\sum_{j = 1}^\Nj g_{y,j} \br{\rho \a}\kj \frac{h\jp-h\jm}{2\dx}.
\label{eq:level_hight_from_sub}
\end{equation}

Potentially, also the pressure term in \eqref{eq:base_model_with_pressure_discretized}
can be improved by back-computing an incompressible subgrid pressure gradient $(\pp_x p)_j$ and then adding the averaging difference ${\an{\ak \pp_x p} - \an \al \an{\pp_x p}}$ as a correction for the non-conservative state integration path. 
Such corrections does not appear to affect the outcome of the presented numerical tests notably 
and so the original \HCU{} pressure term has here been kept unaltered. 
%
\\

\begin{figure}[h!ptb]%
\centering

	\begin{tikzpicture}[align = left,scale = 1.1]
	
		\def\subtextscale{.9}
		
		\coordinate (old v coord) at (3.5,0);
		
		\node[circle,draw] (old V) at (0,0) {old\\ principal\\ state};
		\node[circle,draw] (old v) at (old v coord) {old\\ subgrid\\ state};
		
		
		\node (c3) at (0,-1.9) {$\bullet$};
				
		\node[circle,draw] (new V) at (c3 |- {(0,-3.75)})  {new\\ principal\\ state};
		
		\draw[->] (old V)--(c3) node[left,midway,scale=\subtextscale] {sonic\\flux}; 
		\draw[->,dashed] (old v)--(c3) node[above,midway,scale=\subtextscale](textnode hydraulic) {hydraulic\\flux}; 
		\draw[->] (c3)--(new V) node[left, midway,scale=\subtextscale]{HCU};
		
		\draw[->] (old V) to[bend right=55] node[left,midway,scale = \subtextscale] {pressure\\terms} (new V);
		\draw[->,dashed] (old v) to[bend left=0] node[left,midway,scale = \subtextscale] (textnode source){source\\terms} (new V);
		\draw[->,dashed] (old v) to[bend left=30] node[above,midway,scale = \subtextscale] (textnode lh){level height\\terms} (new V);
	
		\node[rectangle,draw,scale=.9] (P) at (c3 |- {(0,-6.1)}) {$\bullet$ pressure\\$\bullet$ density\\$\bullet$ mixture flow rate};
		\draw[->] (new V)--(P) node[left,midway,scale=\subtextscale] {EOS};

		\node[rectangle,draw] (new v mix) at ( old v coord|- {(0,-3.5)})  {new mixed\\variable $\bv$\\subgrid state};
		\draw[->] (old v)--(new v mix);
		
		\node (c4) at  (new v mix |- P) {$\bullet$};
		\draw[->] (P)--(c4) node[midway,above,scale=\subtextscale]{projection} node[midway,below,scale=\subtextscale]{$\rho\k,\Qm$};
		\draw[->] (new v mix)--(c4);
		\node[circle,draw] (new v) at ( $ (c4) + (1.5,0)  $) {new\\subgrid\\state};
		\draw[->] (c4)--(new v);

		\node[dashed,color = blue!80!black,draw,thick,ellipse,fit= (P.east) (c4)] (ptosub){};
		\node[below,color = blue!80!black,fill opacity=1 ,align=center,scale = .9] at (ptosub.south) {principal coupling:\\principal grid$\rightarrow$ subgrid};
		
		\node[dashed,color = blue!80!black,draw,thick,ellipse,fit= (textnode hydraulic) (textnode source) (textnode lh),scale = .8,xshift=-4mm,yshift=-2mm] (subtop){};
		\node[above,color = blue!80!black,fill opacity=1,scale = .9, anchor=west,xshift=-8mm,yshift=1mm] at (subtop.east) {two-way coupling:\\subgrid $\rightarrow$ principal grid};

	\end{tikzpicture}

\caption{Computational procedure of the dual grid scheme.}%
\label{fig:dual_grid_scheme_procedure}%
\end{figure}
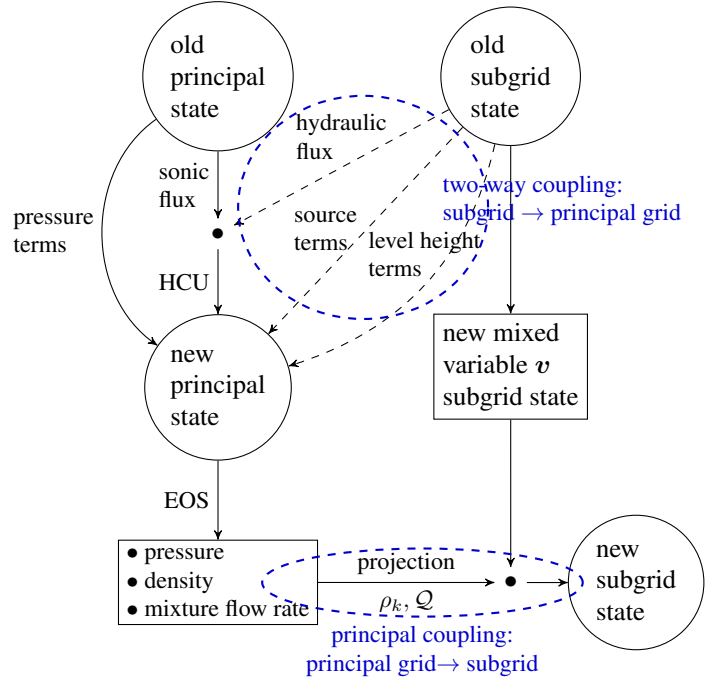

\autoref{fig:dual_grid_scheme_procedure} shows an illustration of the dual grid scheme simulation procedure with the couplings presented in this section.

\section{Properties of the dual grid scheme}

Some features of the dual-grid \HCU{}/Roe scheme are highlighted in this section.
First, in \autoref{sec:vonNeumann}, the linear VKH stability expressions for the differential two-fluid model are presented along with the analogous expressions for the von Neumann stability of the Roe scheme.
It is proven that the proposed Roe scheme will predict the onset of VKH instability \textit{exactly} if allowed a subgrid \CFL{} number equalling one. This property is independent of both the grid resolution and the perturbation wavenumber.

In \autoref{sec:contact_discont} we argue that the dual-grid \HCU{}/Roe scheme maintains the stable properties of the single grid \HCU{} scheme across contact discontinuities. 

Details in this section are not essential for the evaluation that follows.

\subsection{{Kelvin-Helmholtz and von Neumann stability}}
\label{sec:vonNeumann}
{%
The so-called `viscous Kelvin-Helmholtz' (VKH) stability analysis is well known in literature (\egg, \cite{Barnea_VKH_and_IKH,Lin_Henratty_VKH},) 
and can be related to discrete representations of the same model with similar stability expressions \citep{Stewart_stability_well_illposed,Fullmer_VKH_von_Neumann}.
Derivation and validation of these analyses in the form presented here can be found in, \egg,  \citep{Akselsen_PhD}.

The dispersion equation for the incompressible two-fluid model \eqref{eq:base_model} reads}
\begin{align}
	&\sqbrac{\br{ \frac{\rho\l}{\al}+\frac{\rho\g}{\ag}}(\lambda_+-c)(\lambda_--c)}\delta_x  \nonumber \\
	&\qquad-\sqbrac{\pdiff{s}{\al} - \frac{u\l-c}{\al}\pdiff{s}{u\l} + \frac{u\g-c}{\ag}\pdiff{s}{u\g}}=0
\label{eq:disp_eq}
\end{align}
%
where
$c$ is the (complex) phase velocity of the perturbation wave. 
Spatial differentiation manifests as
$\delta_x = \imunit k$ in a differential model and as $\delta_x = \br{1-\exp(-\imunit k\dx)}/\dx$ in a discrete model representation using the Roe scheme, assuming vigorous flow towards the right.%
\footnote{
Both characteristics will be of the same sign 
close to and beyond the stability limit.
The Roe scheme is then described by \eqref{eq:Roe_supercritical}.
A general expression is presented in \cite{Akselsen_PhD}.}
$k$ is here the perturbation wavenumber.
The dispersion equation \eqref{eq:disp_eq} is an easily solvable second order equation for $c$. 
$c$ is the actual complex perturbation wave velocity in the case of the differential model, \ie,
\begin{align*}
	\text{model \eqref{eq:base_model}:}\qquad \a\of{x,t}&\sim \eEuler^{\imunit k (x-ct)}. 
\intertext{%
{%
A positive imaginary value of $c$ corresponds to wave growth in the differential model and hence VKH instability.
The actual complex wave velocity in a discrete representation, $c\_d$,  
}%
}%
\text{scheme \eqref{eq:Roe_scheme}:}\qquad \alj\n &\sim \eEuler^{\imunit k (x_j-c\_d t_n)}, 
\end{align*}
{%
will depend on the time integration. 
In the case of simple Euler and Crank-Nicolson time integration schemes we may write
}%
\begin{align}
	 c\_d &=  \frac{\imunit}{k\dt} \ln\br{\frac{1-(1-r)c\dt\delta_x}{1+rc\dt\delta_x}}, \label{eq:vonNeumann:c_explicit}
\end{align}
with $r=0$ for forwards Euler, $r=1$ for backwards Euler and $r=1/2$ for Crank-Nicolson time integration.

{
From this we can demonstrate, possibly for the first time, the fact that the Roe scheme (and indeed the simple upwind $\bf\jph = \bf_j$) will predict the onset of VKH instability exactly if the \CFL{} number equals $1.0$. The proof goes as follows:}

{
First, the VKH condition for wave growth in the differential model can be written
}%
\[
	(\lambda_+-c)(\lambda_--c)<0,
\]
{%
with equality indicating a neutrally stable state. This holds for all wavenumbers and can be inferred form \eqref{eq:disp_eq} by imposing that $c$ is real. 
Still assuming flow towards the right, this condition entails that $\lambda_+=c$ at the state of neutral VKH stability. 
A \CFL{} number equalling one locally thus implies
}%
\[
	\frac{\dx}{\dt} = \max|\lambda_\pm| = \lambda_+ = c.
\]
{%
Thus inserting $\dx = c \dt$ into the explicit forward Euler time integration ($r=0$ in \eqref{eq:vonNeumann:c_explicit}) we recover
$c = c\_d$. }

Consider next the the differential model. A root $c$ of the dispersion equation \eqref{eq:disp_eq} is real at neutral stability.
$c$ is then a root of each of the square bracket terms individually because $\delta_x=\imunit k $ is purely imaginary in the differential model.
The same real $c$ is also a root of the discrete dispersion equation, even though $\delta_x$ is now some complex value.
As just shown, this real $c$ equals $c\_d$ at $\CFL{}=1.0$ and therefore the perturbation is neutrally stable also in the discrete representation. 
\\

In other words, the explicit Roe scheme detects the onset of VKH instability \textit{exactly} if allowed a \CFL{} number equalling one, regardless of the wavenumber and grid resolution. 
The neutrally stable perturbation has then also got the correct phase velocity.
This result is analogous to the well-known fact that the exact solution of the linear advection equation is obtained with an upwind scheme and $\CFL{}=1.0$.

At states \textit{close} to the neutrally stable one, such as the state shown later in \autoref{fig:exp:roll_wave:disc_stab}, the accuracy will be good with $\CFL{}=1.0$, but not exact nor wavenumber independent.

\subsection{The contact discontinuity}
\label{sec:contact_discont}

The \HCU{} scheme \citep{Evje_HCU_schemes} was designed to resist numerical errors across contact discontinuities. 
This property is maintained in the dual-grid scheme, as argued below.\\

Consider the following contact discontinuity, analogous to (32) in \citep{Evje_HCU_schemes}:
\begin{equation}
\left\{
\begin{gathered}
u\_{\ell,L} = u\_{g,L} = u\_{\ell,R} = u\_{g,R} = u
\\
\a\_{\ell,L} \neq \a\_{\ell,R}\\
p\_L = p\_R
\\
g \rightarrow 0
\\
s \equiv 0
\end{gathered}
\right\}
\label{eq:contact_discont}
\end{equation}
The two-fluid model \eqref{eq:base_model_with_pressure} reduces to 
\begin{equation}
	\pp_t \al + \pp_x u \al = 0
\label{eq:contact_discont_sol}
\end{equation}
under these conditions, and the problem solution is simply a uniform advection of the initial state, \ie,
$
\al\of{x,t} = \al\of{x-u t, 0}.
$


Consider next the Roe scheme \eqref{eq:Roe_scheme} with the two-way coupling of \autoref{sec:dual_grid_scheme:couplings}.
Note first that $\symkappa$ and the eigenvalues $\lambda^\pm$ in the Roe scheme \eqref{eq:Roe_sol_f} 
will in the contact discontinuity \eqref{eq:contact_discont} reduce to 
$
\symkappa \rightarrow 0$ and $\lambda^+,\lambda^- \rightarrow u, 
$
and so
the Roe matrix \eqref{eq:Lambda_LL} reduces to
$
\tfrac12 |\Roe|\jph \rightarrow |u|\I.
$
The Roe scheme fluxes \eqref{eq:Roe_sol_f} then reduce to
\begin{align*}
\bf\jph \rightarrow \begin{pmatrix}
	\h\br{\a\_{\ell,R} + \a\_{\ell,L} }u - \h\br{\a\_{\ell,R} - \a\_{\ell,L} }|u|\\
	\h\br{\rho\l-\rho\g} u^2
\end{pmatrix}.
\end{align*}
Also the simple upwind scheme \eqref{eq:f_simple_upwind} will reduce to this same expression.

The mixture flow rate $\Qm_J = \frac{I\lJ}{\rho\lJ}+\frac{I\gJ}{\rho\gJ} $ reduces to the constant $\area\, u$ in the contact \eqref{eq:contact_discont}, and
the sub-to-principal grid couplings \eqref{eq:F_U_subgrid} and \eqref{eq:UI_subgrid} then reduce to
%
%
\begin{align*}
\F_{\phaseindex,J+\h}\UU & \rightarrow
\begin{cases}
\rho\k \a_{\phaseindex,\rm L} u& \text{if}\quad u\geq0,\\
\rho\k \a_{\phaseindex,\rm R} u& \text{if}\quad u<0,
\end{cases} 
\end{align*}
and
\begin{equation*}
\br{UI}\kJph \rightarrow u \F_{\phaseindex,J+\h}\UU.
\end{equation*}
These are the simple upwind fluxes for the linear advection problem \eqref{eq:contact_discont_sol}
and so upwind fluxes $\F\k\UU$ are consistent with the contact discontinuity solution.



Note further that
 all other variables in \eqref{eq:flux_splitting_result} pertain to the principal grid, computed as in the original \HCU{} scheme. 
The other subgrid couplings, \eqref{eq:S_from_sub}, \eqref{eq:level_hight_from_sub} (and \eqref{eq:non_stratified_term} presented later), reduce to zero in the contact \eqref{eq:contact_discont}.
%
%
The remarks, propositions and proofs presented in \citep{Evje_HCU_schemes} therefore also hold for the dual grid scheme, meaning that the central mass fluxes $\F\k\CC$ are `pressure preserving' and that 
flows which are initially uniform in pressure and velocity will remain so in time.
Numerical pressure disturbances are thus avoided across the contact discontinuity.
%
%
%
\\

\begin{figure}%
\hspace{-.1\columnwidth}
\begin{subfigure}{1.2\columnwidth}
\begin{minipage}{.49\columnwidth}\centering
\includegraphics[width=\columnwidth]{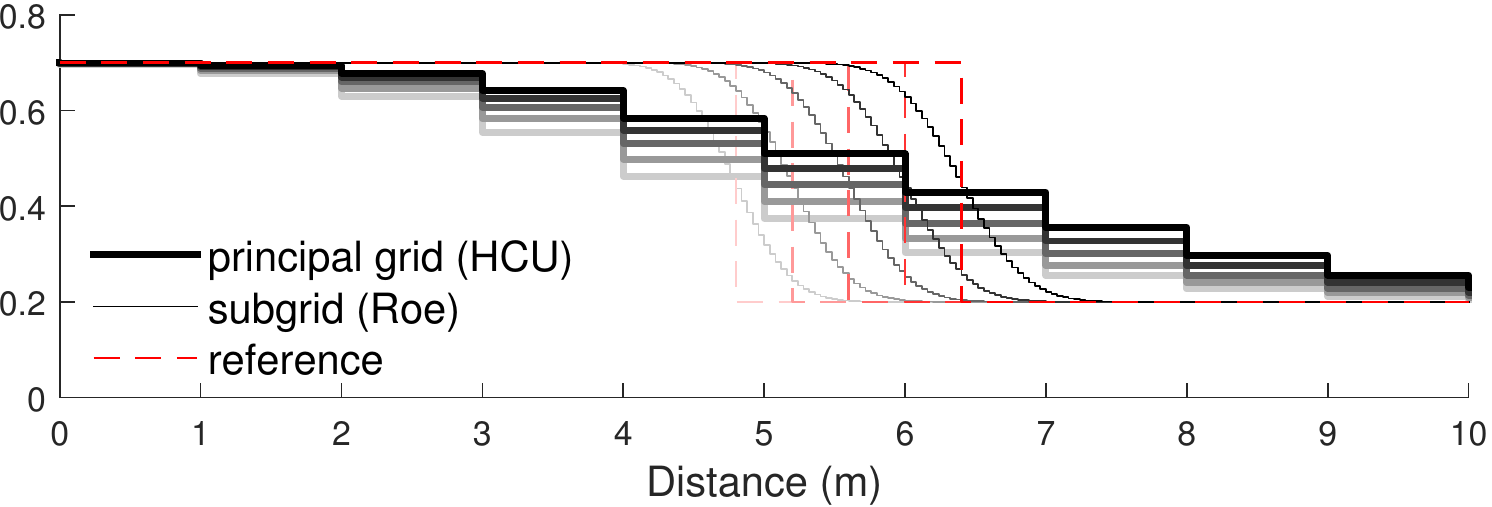}%
\\ {\small One-way coupling\\ (\autoref{sec:dual_grid_scheme:integral_to_sub} only.)}
\end{minipage}
\begin{minipage}{.49\columnwidth}\centering
\includegraphics[width=\columnwidth]{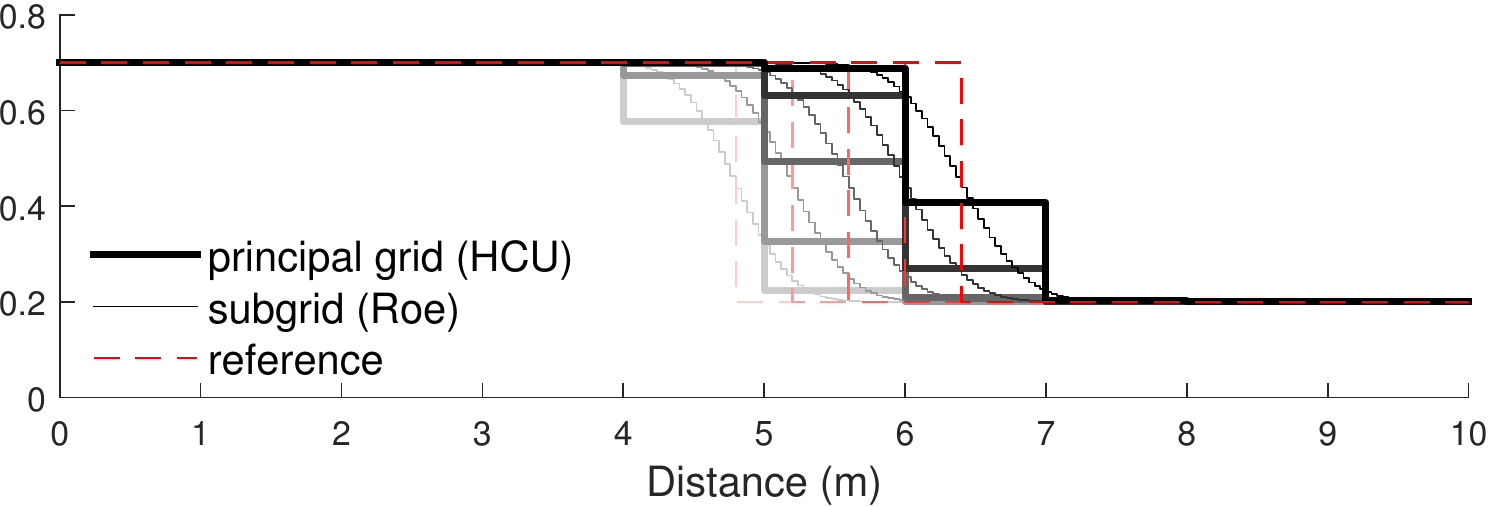}%
\\ {\small Two-way coupling\\ (\autoref{sec:dual_grid_scheme:integral_to_sub} and \ref{sec:dual_grid_scheme:sub_to_integral}.)}
\end{minipage}
\caption{Liquid fraction.}
\end{subfigure}
\\ 
\begin{subfigure}{\columnwidth}
\centering
\includegraphics[width=.675\columnwidth]{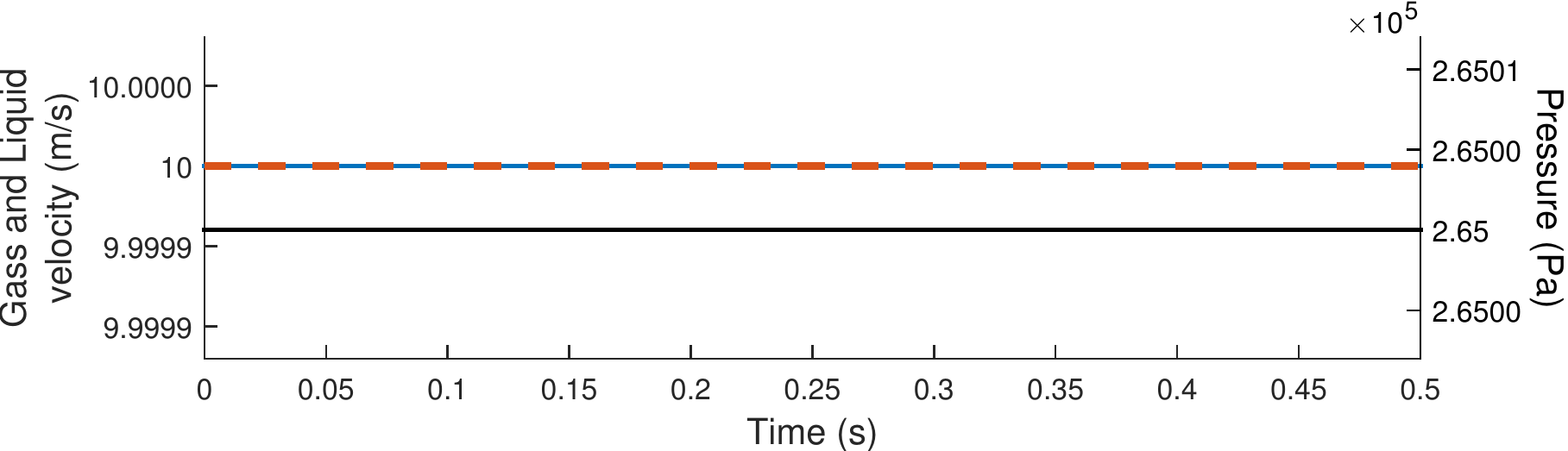}%
\caption{Time trace with the two-way coupling at fixed location as contact passes.}
\end{subfigure}
\caption{Contact discontinuity -- verification of undisturbed contact.\\
($u\_{g,L}=u\_{g,R}=u\_{\ell,L}=u\_{\ell,R} = \unitfrac[10]ms$, $p\_L=p_R = \unit[265\,000]{Pa}$, $s=0, g\rightarrow0$.)}%
\label{fig:contact_discont}%
\end{figure}

As a verification, \autoref{fig:contact_discont} shows a contact discontinuity and a plot monitoring the pressure and phase velocities; the otherwise uniform state is undisturbed by the passing contact.

\section{Numerical tests}
\label{sec:exp}

\subsection{Benchmark tests (dispersive flow)}
\label{sec:exp:dispersive_tests}

We will begin with two basic benchmark test cases found in \citep{Evje_HCU_schemes,Flaatten_common_two_fluid}. 
Both the one-way coupled and two-way coupled variants of the dual grid scheme are tested.
If only one-way coupled, the \HCU{} scheme in the principal grid is unadjusted and unaware of the subgrid.
Principal grid results are then identical to that presented in \citep{Evje_HCU_schemes}. 
The coupling proposals in \autoref{sec:dual_grid_scheme:sub_to_integral} are applied in the two-way coupled grid arrangement. 
\\

%
The flow in \citep{Evje_HCU_schemes} 
is dispersed. 
For comparability, we convert the stratified model described thus far into one for dispersed flow by replacing the hydrostatic pressure term $g_y m\k  \pp_x h$  with $\Delta p\, \pp_x \ak$ in the momentum equation~\eqref{eq:base_model_with_pressure:mom}. 
The new interface pressure term
\begin{equation}
	\Delta p = \frac\sigma\area \frac{\al \ag \rho\l\rho\g}{\rho\g\al + \rho\l\ag} \br{u\g-u\l}^2
	\label{eq:non_stratified_term}
\end{equation}
has a physical interpretation for bubbly flows only, but ensures model hyperbolicity if $\sigma>1$ (see \eqref{eq:symkappa}.) 
 $\sigma = 1.2$ is used in these simulations.
\added{
The Roe subgrid scheme \eqref{eq:Roe_sol_f} must also be modified for dispersed flow. This is done by replacing $\symkappa$ in the eigenvalues \eqref{eq:eigenvalues} with
}
\[
\symkappa = \sqrt{(\sigma-1)\frac{\rho\l \rho\g}{\al \ag}(u\g-u\l)^2}
\]
\added{%
and otherwise computing the Roe matrix as before.
}
$\Nj{}=40$ subgrid cells are applied per principal cell in these tests.

Fluid properties in the equations of state \eqref{eq:EOS_linear} are
\begin{align*}
\rho_{\ell,0} &= \unitfrac[1000]{kg}{m^3},
&
\rho_{\mr g,0} &= \unitfrac[0]{kg}{m^3},
&
p\_{0} &= \unit[10^5]{Pa},
\\
\drholdpEOS &= \unitfrac[10^{-6}]{m^2}{s^2},
&
\drhogdpEOS &= \unitfrac[10^{-5}]{m^2}{s^2}. &&
\end{align*}
No wall or interfacial friction is present in these problems.

\subsubsection{Shock tube problem}
\label{sec:exp:shock_tube}
\begin{figure*}[h!ptb]%
\hspace{-.15\linewidth}
\includegraphics[width=1.3\linewidth]{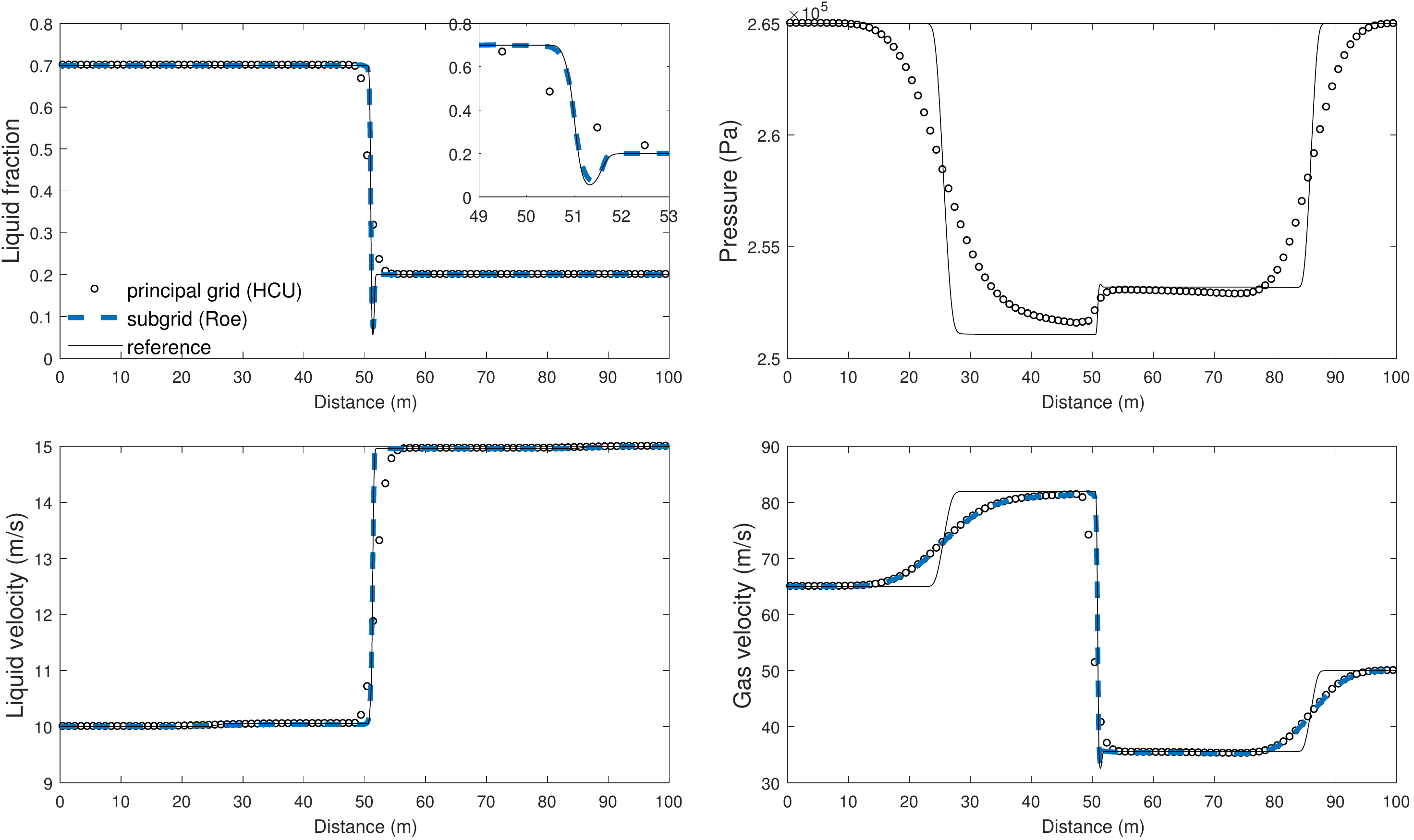}%
\caption{Modified LRV shock tube problem, cf.\ Fig.~4 in \citep{Evje_HCU_schemes}. 
$t = \unit[0.1]s$, 
$\dX/\dt=\unitfrac[750]ms$. 
One-way coupling (suggestions in \autoref{sec:dual_grid_scheme:sub_to_integral} are not applied, \ie, the original \HCU{} scheme is active in the principal grid.)}%
\label{fig:exp:Flatten_noncoupled}%
\end{figure*}

\begin{figure*}[h!ptb]%
\hspace{-.15\linewidth}
\includegraphics[width=1.3\linewidth]{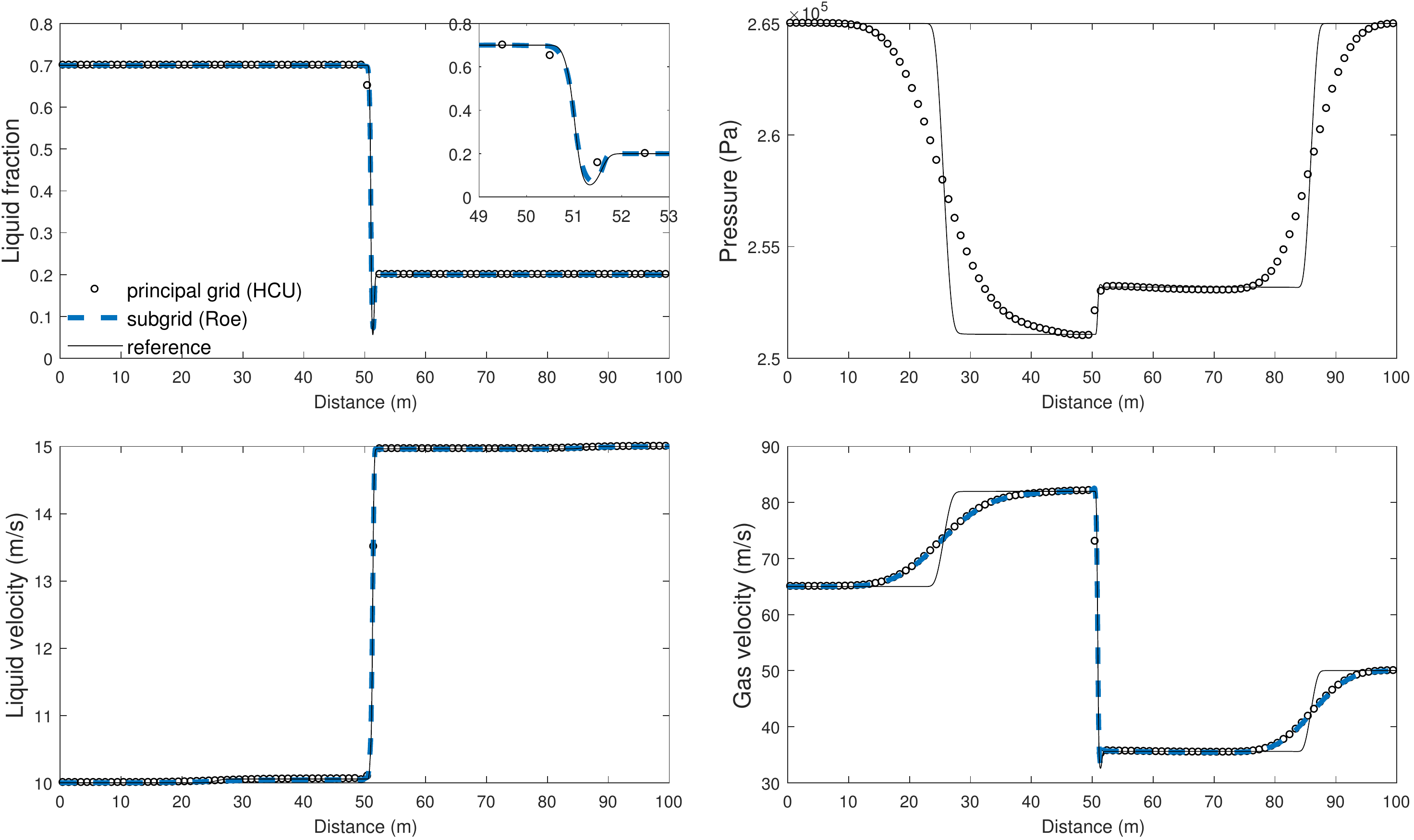}%
\caption{Modified LRV shock tube problem, cf.\ Fig.~4 in \citep{Evje_HCU_schemes}.
$t = \unit[0.1]s$, 
$\dX/\dt=\unitfrac[750]ms$.
Two-way coupling (suggestions in \autoref{sec:dual_grid_scheme:sub_to_integral} applied.)}%
\label{fig:exp:Flatten_coupled}%
\end{figure*}

\autoref{fig:exp:Flatten_noncoupled} and \autoref{fig:exp:Flatten_coupled} show the modified large relative velocity (LRV) shock tube problem studied in \citep{Evje_HCU_schemes,Flaatten_common_two_fluid}. Left and right initial states are
\begin{align*}
\begin{bmatrix}
p\\ \al/\!\area\\ u\g \\ u\l	
\end{bmatrix}\_L
\!\!&=
\begin{bmatrix}
\unit[265\,000]{Pa}\\
0.7\\
\unitfrac[65]ms\\
\unitfrac[10]ms
\end{bmatrix},
&
\begin{bmatrix}
p\\ \al/\!\area\\ u\g \\ u\l	
\end{bmatrix}\_R
\!\!&=
\begin{bmatrix}
\unit[265\,000]{Pa}\\
0.1\\
\unitfrac[50]ms\\
\unitfrac[15]ms
\end{bmatrix}.
\end{align*}
%

$\NJ{}=100$ principal cells are used with a time step 
$\dX/\dt=\unitfrac[750]ms$ ($\dt \approx \unit[0.013]s$.)
\deleted{
The Roe scheme (3.14) is only applicable to the stratified two-fluid model; the simple upwind scheme (3.21) is instead employed for the subgrid model (3.10) with the alteration (5.1).
}
The reference is computed with the single grid \HCU{} scheme using 10\,000 grid cells.
\\

\autoref{fig:exp:Flatten_noncoupled} shows a one-way coupled simulation where only the couplings in \autoref{sec:dual_grid_scheme:integral_to_sub} are applied,
and the principal grid results are therefore identical to that in Figure~4. of \citep{Evje_HCU_schemes}.
Flow details attributed to hydraulic waves 
are seen to be accurately recovered in the subgrid.
Mass flux consistency is not maintained between grids in this one-way coupled simulation and so volume fractions are not within proximity;
numerical diffusion is stronger in the principal grid than in the subgrid.
\\

The full two-way coupling is adopted in \autoref{fig:exp:Flatten_coupled}, showing the same simulation case. 
Mass proximity is here ensured through the flux \eqref{eq:F_U_subgrid:liq}, errors from liquid compressibility being insignificantly small.  
Subgrid fluxes \eqref{eq:F_U_subgrid} and \eqref{eq:UI_subgrid} also provide high accuracy and 
tight
proximity in the fluid velocities. 
Pressure is not computed in the subgrid, but the pressure prediction is improved indirectly via the other equation terms.
Term-by-term testing indicates that it is the computation of \eqref{eq:non_stratified_term} from the subgrid data that contributed the most to this improved pressure prediction --  
it is here computed analogously to \eqref{eq:level_hight_from_sub}, namely
\[
\br{\Delta P\, \pp_x \Al}_J = \frac1\Nj \sum_{j=1}^\Nj (\Delta p)_j \frac{\a\ljp-\a\ljm}{2\dx}.
\] 
Interpolation errors form the initially discontinuous mixture flow rate 
are present in the initial stages of the simulation but quickly die out as pressure diffusion kicks in.


\subsubsection{\added{Water faucet problem}}
\label{sec:exp:water_faucet}

\added{
The water faucet benchmark problem by \cite{Ransom_1987_water_faucet} is frequently studied, also in \citep{Evje_HCU_schemes} and the references therein. 
The uniform initial state of this problem is}
\begin{equation*}
\begin{bmatrix}
p\\ \al/\!\area\\ u\g \\ u\l	
\end{bmatrix}
=
\begin{bmatrix}
\unit[10^{5}]{Pa}\\
0.8\\
\unitfrac[0]ms\\
\unitfrac[10]ms
\end{bmatrix}
\end{equation*}
\added{%
within a \unit[12]m long vertical pipe. 
Corresponding boundary conditions, $\al/\!\area=0.8$, $u\g =\unitfrac[0]ms$ and $u\l =\unitfrac[10]ms$ at the inlet and $p=\unit[10^{5}]{Pa}$ at the outlet, apply. 
Gravity causes fluid present in the pipe to accelerate, creating a downward propagating shock.  
\\}

\begin{figure*}[h!ptb]%
\hspace{-.15\linewidth}
\includegraphics[width=1.3\linewidth]{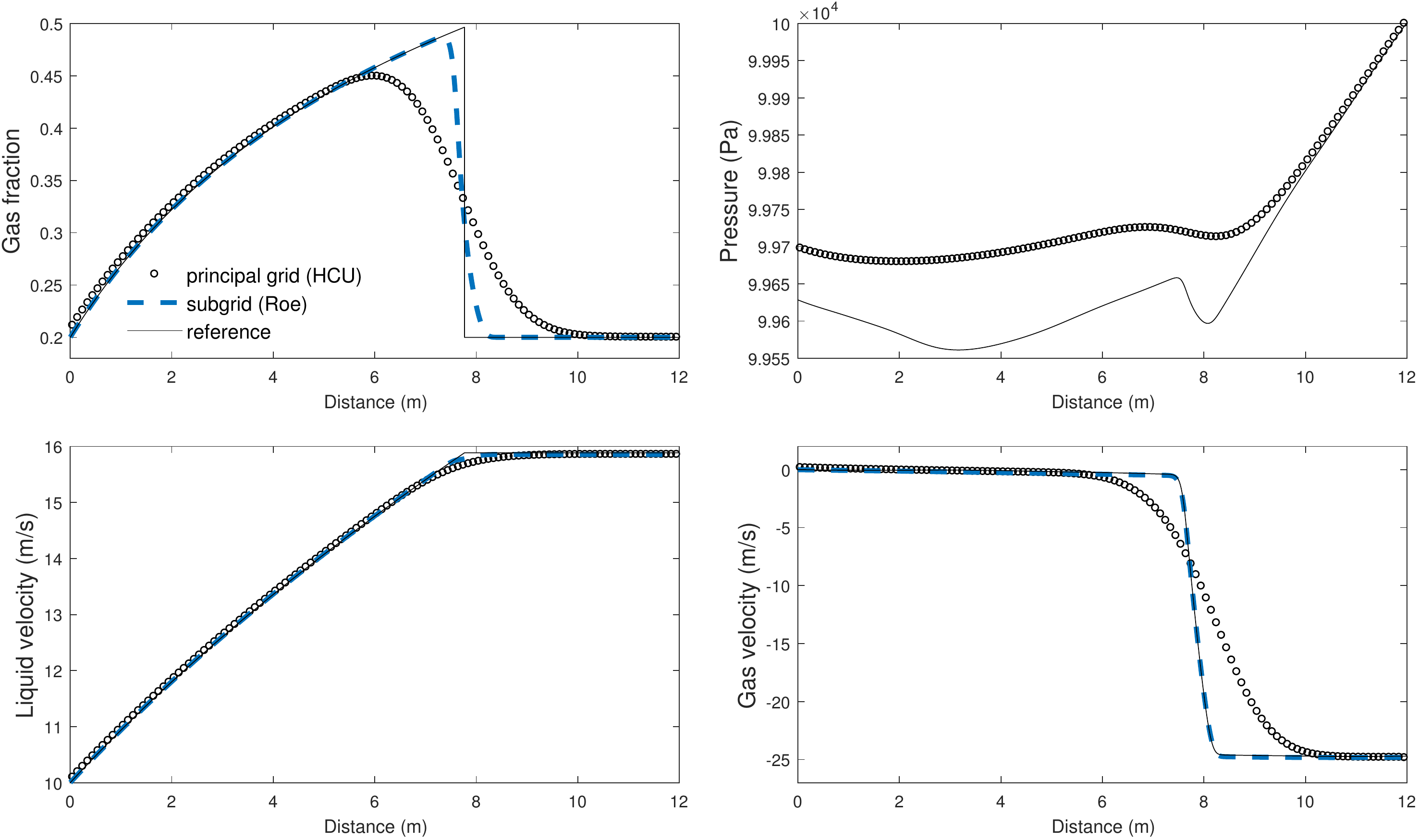}%
\caption{Water faucet problem, cf.\ Fig.~8 in \citep{Evje_HCU_schemes}.  $t = \unit[0.6]s$, 
$\dX/\dt = \unitfrac[10^3]ms$. 
One-way coupling (suggestions in \autoref{sec:dual_grid_scheme:sub_to_integral} are not applied, \ie, the original \HCU{} scheme is active in the principal grid.)
}%
\label{fig:exp:water_faucet_noncoupled}%
\end{figure*}

\begin{figure*}[h!ptb]%
\hspace{-.15\linewidth}
\includegraphics[width=1.3\linewidth]{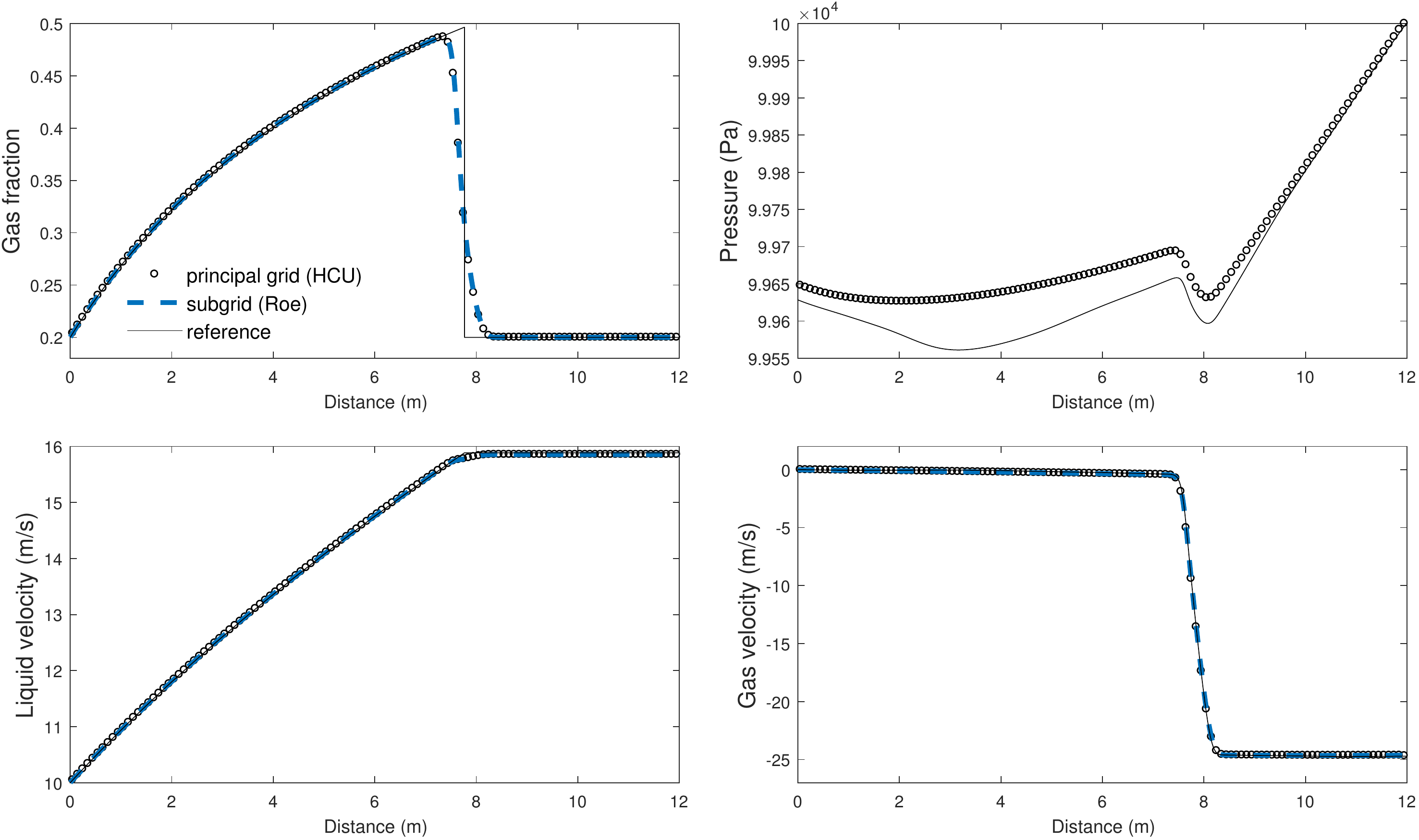}
\caption{Water faucet problem, cf.\ Fig.~8 in \citep{Evje_HCU_schemes}. 
$t = \unit[0.6]s$, 
$\dX/\dt = \unitfrac[10^3]ms$. 
Two-way coupling (suggestions in \autoref{sec:dual_grid_scheme:sub_to_integral} applied.)
}%
\label{fig:exp:water_faucet_coupled}%
\end{figure*}

\added{%
An approximate analytical solution quoted in \egg\ \citep{Evje_HCU_schemes} can be derived for this problem.
\autoref{fig:exp:water_faucet_noncoupled}--\ref{fig:exp:water_faucet_coupled} compares the one-way and two-way coupled schemes, respectively, to this analytical solution. 
$NJ=120$ principal grid cells and a time step
$\dX/\dt = \unitfrac[10^3]ms$ ($\dt=\unit[10^{\text{-}4}]{s}$) are used.
Pressure and gas velocity references, not available from the analytical expression, has here been computed with the single grid \HCU{} scheme using $12\,000$ grid cells. 
This is again the same benchmark setup found in \citep{Evje_HCU_schemes}, cf.\ Figure~8.
}

\added{
	As before we observe significant improvements to the predicted solution due to the subgrid. 
	Pressure predictions notably improve with the two-wave coupling and no signs of grid-related disturbances can be detected. 
}


\subsection{Large cases problems (stratified flow)}
\label{sec:exp:stratified_tests}

The two final test cases are directed towards engineering-type problems.
We use the parameters
\begin{align*}
\rho_{\ell,0} &= \unitfrac[1000]{kg}{m^3},
&
\rho_{\mr g,0} &= \unitfrac[50]{kg}{m^3},
\quad
p\_{0} = \unit[8\Ep5]{Pa},
\\
\drholdpEOS \! &= \unitfrac[0]{m^2}{s^2},
&
\drhogdpEOS \! &= \unitfrac[7.77\Em 5]{m^2}{s^2},
\end{align*}
for the equations of state \eqref{eq:EOS_linear}, 
and the fluid and pipe properties\\
\begin{tabular}[b]{llrl}
liquid viscosity&	$\mu\l$ 	& $1.00\Em3$&		$\unit{Pa\:s}$\\ 
gas viscosity&		$\mu\g$ 	& $1.61\Em5$&		$\unit{Pa\:s}$ \\
internal pipe diameter&			& $0.1$&			$\unit[]m$	\\
wall roughness&							& $2\Em5$& 		$\unit[]m$ \\
pipe inclination& 			$\theta$	& $0\degree$&		$-$\\
\end{tabular}.
\\
These properties correspond to the experimental and numerical setup used in \citep{Holmaas_roll_wave_model, George_PhD,Akselsen_char_Roe}.
The friction closures $\tau\k$ and $\tau\_i$ are from the Biberg friction model as presented in \citep{Biberg_friction_duct_2007}, also described in the other references just mentioned.

\subsubsection{Surge wave case}
\label{sec:exp:surge_wave}

A  $\unit[100]m$ long horizontal pipeline is simulated with an outlet pressure initially at $8$ bara.
A constant liquid mass flux of $\unitfrac[1.5]{kg}s$ and a constant gas mass flux of $\unitfrac[1.0]{kg}s$ are supplied at the inlet. 
Liquid fraction at the inlet is $0.19$, giving a fairly smooth transition into the test section.
The initial state in the pipeline corresponds to the superficial velocities $\al u\l/\area \approx \unitfrac[0.1]ms$, $\ag u\g/\area \approx \unitfrac[3.1]ms$, which does \textit{not} match the inlet conditions. 
A surge wave starting from the inlet at $t=0$ results.
Further, at $t=\unit[30]s$, we instantaneously reduce the outlet pressure to 7.5 bara, creating a pressure wave moving towards the inlet. This pressure wave will interact with the inlet and the surge wave. %

\begin{figure*}
\begin{minipage}{\linewidth}
\vspace{-25mm}
\hspace{-.21\linewidth}
\includegraphics[width=1.45\linewidth]{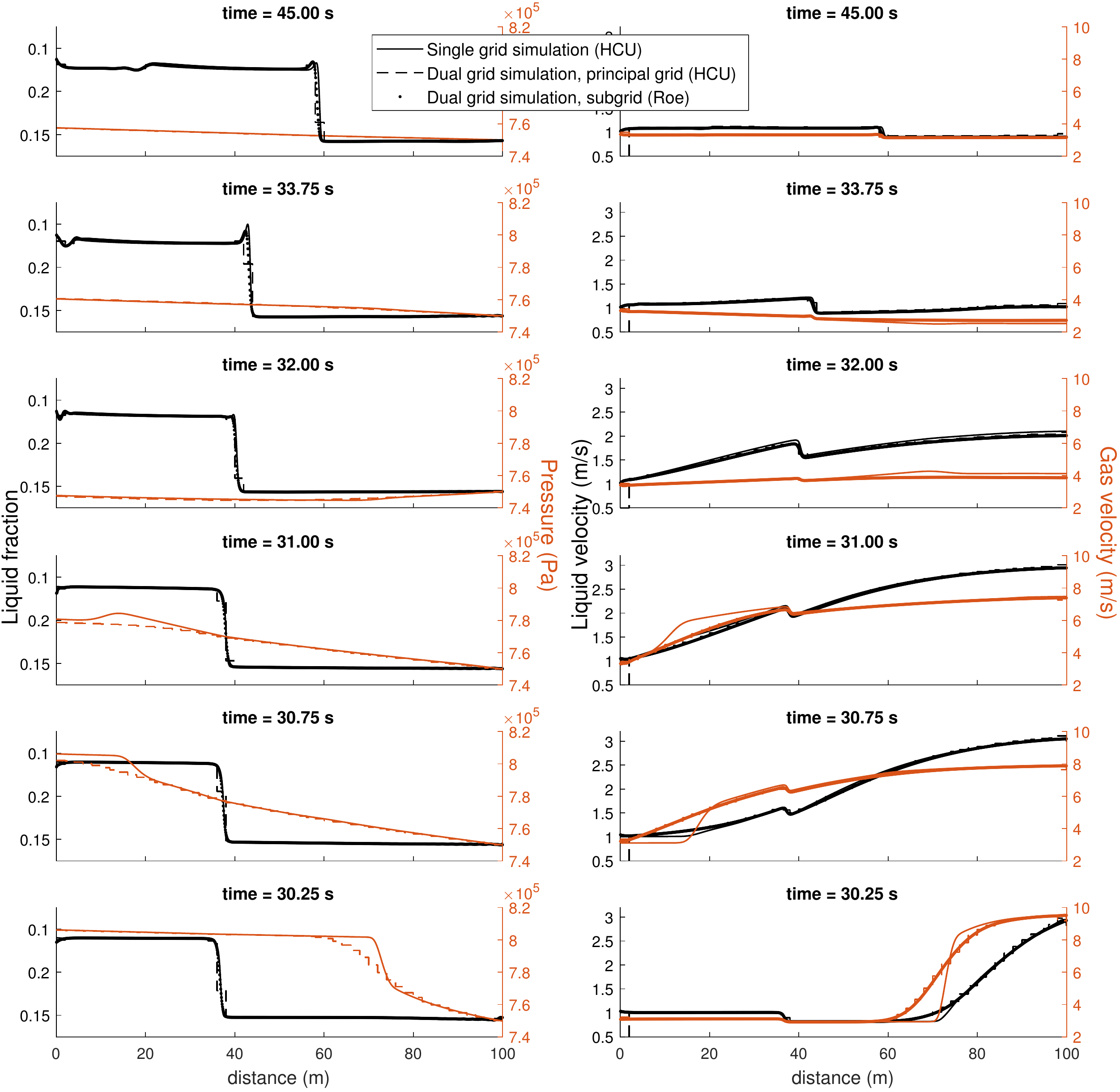}%
\subcaption{$\Nj = 25$ subgrid cells per principal cell. }%
\label{fig:exp:surge_wave:HD1}%
\end{minipage}
\begin{minipage}[c]{.5\linewidth}
\hspace{-.3\linewidth}
\includegraphics[width=1.3\linewidth]{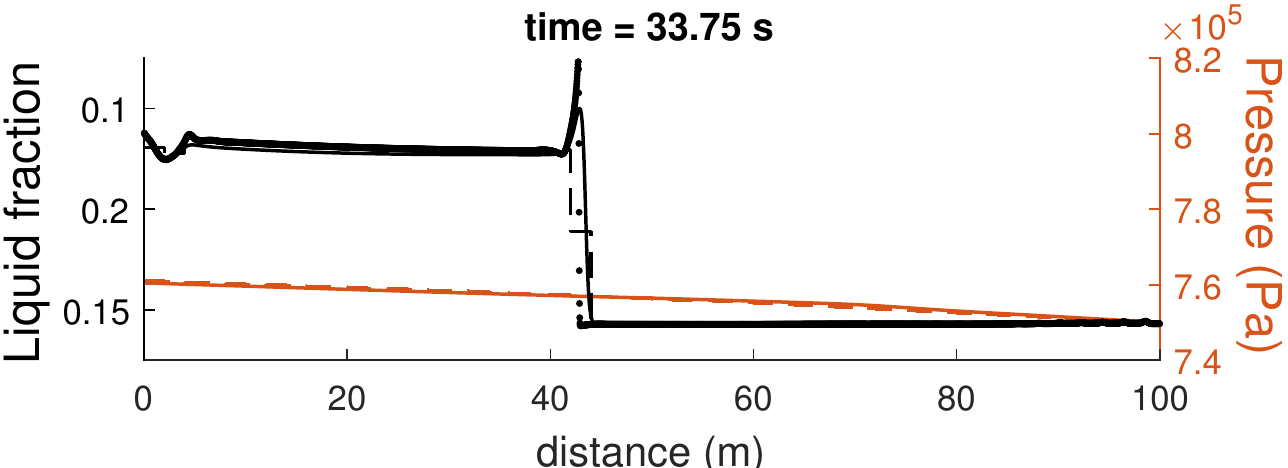}%
\end{minipage}
\begin{minipage}[c]{.5\linewidth}
\subcaption{$\Nj = 100$ subgrid cells per principal cell. }%
\label{fig:exp:surge_wave:HD2}%
\end{minipage}
\caption{Surge and pressure waves. A single grid and a dual grid simulation plotted in the same axes.
\\$\NJ = 50$ and $\dt = \unit[0.01]s$ in the dual grid simulation.
\\$\NJ = 2\,000$ and $\dt = \unit[0.00025]s$ in the single grid simulation.}
\label{fig:exp:surge_wave:HD}%
\end{figure*}

\autoref{fig:exp:surge_wave:HD1} shows snapshots of two such surge wave simulations, one with the single grid \HCU{} scheme (functioning as a reference) and the other with the dual grid \HCU{}/Roe scheme. 
2\,000 grid cells are used in the single grid \HCU{} scheme in order for the surge and pressure waves to maintain fairly sharp fronts.
The time step is  $\dX/\dt = \unitfrac[200]ms$ ($\dt = \unit[2.5\Em4]s$,) which is close to the numerical stability limit.

In the dual grid \HCU{}/Roe scheme $\NJ = 50$ principal cells are applied with $\Nj = 25$ 
subgrid cells per principal cell. 
The time step is also here restricted by the sonic speed such that $\dX/\dt = \unitfrac[200]ms$ ($\dt=\unit[0.01]s$.)

The time of each snapshot increases (unevenly) through the panels, showing the pressure wave as it passes over the surge wave front and bounces back off of the inlet. 
This pressure wave bounces back and forth across the pipeline a couple of times before dying out. 
The surge wave reacts relatively slowly to the change in pressure, responding with a temporary peak forming at the front.
New hydraulic waves are also created at the inlet in tune with the bouncing  pressure wave.
\\

Compared to the single grid simulation, the dual grid shows a more diffusive pressure wave but nearly no differences in the surge wave. 
Subgrid and principal grid data remain within close proximity;
the subgrid data points form lines through their respective principal grid stair plots (except for pressure, which is not computed in the subgrid.)
Because of the volume flux consistency, liquid fractions in both grids are always perfectly synchronized without the orientation of one grid disturbing the other.
\\

In terms of efficiency, the single grid requires $8\,000\,000$ 
\HCU{} cell computations per simulated second. 
In comparison, the dual grid simulation requires $5\,000$ \HCU{} computations and $125\,000$   
Roe cell computations per simulated second.
The equation of state is computed in the \HCU{} cells only.

Resolution in the dual grid scheme was here chosen to give a phase fraction solution similar to the single grid scheme. 
A reasonable resolution of the sonic wave was also acquired.
Indeed, because it is the sonic speeds which limit the time steps we may increase the number of subgrid cells even further at only linearly increasing computational cost. 
A single panel snapshot where the subgrid resolution is $\Nj = 100$ is shown in \autoref{fig:exp:surge_wave:HD2}.
Accuracy is further improved by the fact that the Roe scheme becomes more accurate as the CFL number \eqref{eq:Roe_CFL} approaches unity. 
\\

Let's go even further.
We now use only $\NJ=10$ principal cells
and a sufficient number of subgrid cells for the time step to be restricted by the hydraulic $\CFL{}=1.0$ limit in \eqref{eq:Roe_CFL}, rather than by
the sonic time scale.
This warrants the choice of $\Nj = 125$ subgrid cells per integral cell, yielding a $\CFL{}=1.0$ time step of $\dx/\dt\approx\unitfrac[1.38]ms$ ($\dt \approx \unit[0.058]s$.) 

\begin{figure*}
\hspace{-.25\linewidth}
\includegraphics[width=1.5\linewidth]{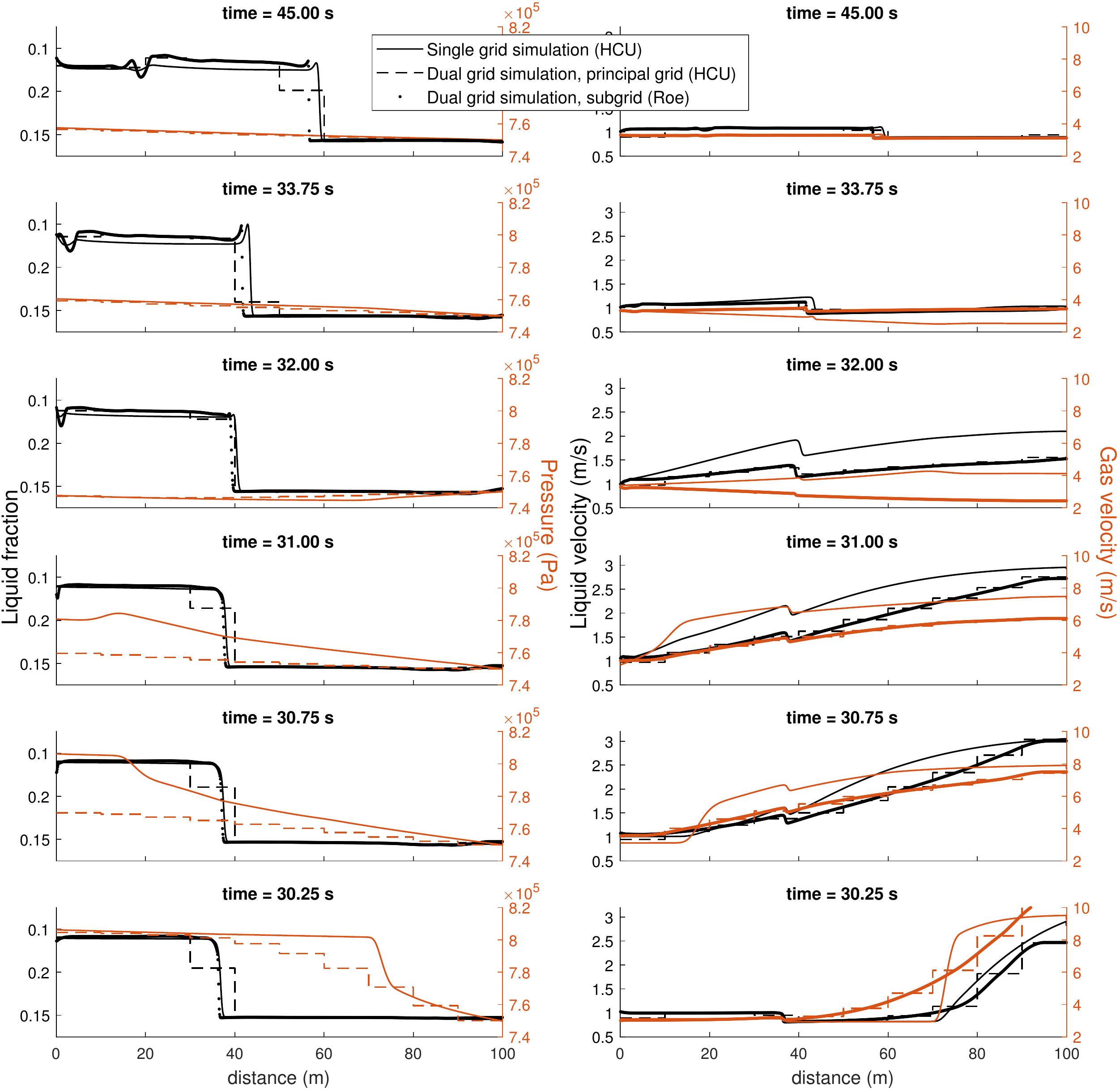}%
\caption{Surge and pressure waves. A single grid and a dual grid simulation plotted in the same axes.
\\$\NJ = 10$, $\Nj = 125$ and $\dt\approx\unit[0.058]s$ ($\CFL{}=1.0$) in the dual grid simulation.
\\$\NJ = 2\,000$ and $\dt = \unit[0.00025]s$ in the single grid simulation.}
\label{fig:exp:surge_wave:LD}%
\end{figure*}

\autoref{fig:exp:surge_wave:LD} shows the resulting simulations compared with the same single grid reference. 
The pressure wave front is now more or less lost to numerical diffusion, but the switch from one steady pressure condition to another is reasonably accurate.
Details of the volume fractions are captured and highly resolved. 
With the Roe CFL number at unity, a very sharp surge wave front is observed some time after the pressure drop.
The inlet wave is also generated appropriately. Its wavelength resembles the wavelength of the reference, suggesting that errors in the sonic velocities are not severe, despite the diffusiveness of the pressure wave. 
Because of the high CFL number, these inlet waves do not die away due to numerical diffusion.

Some discrepancies are seen with the liquid fraction late in the simulation. 
Most notably, the surge wave front 
lags slightly behind the single grid front in the later panels, after the pressure drop.
Liquid fraction in between the inlet wave and surge wave front is consequently higher.
The liquid fraction after the inlet wave is again accurate, suggesting that this discrepancy is caused by the numerical dissipation of the pressure wave.
The boundary cells may also have contributed to this discrepancy.
\\

Again we consider the increase in computational efficiency. The dual grid simulation now required about $175$ \HCU{} cell computations and $22\,000$ 
Roe cell computations per simulated second. 
Assuming the cost of a Roe and a $\HCU$ cell computation are comparable, the single grid \HCU{} simulation is more than $350$ times as computationally expensive as the latter dual grid simulation.

\subsubsection{Roll-wave case}
\label{sec:exp:roll_wave}
Let us finally consider flow in the roll-wave regime. 
This type of flow is on the whole unaffected by compressibility. 
All the same we want the ability to simulate it with a compressible model in order to account for expansion effects, slug formations, changing boundary conditions, etc..
Johnson's high-pressure experimental campaign \citep{George_PhD} is chosen for the simulation setup. 
\\

For briefness, only a single high gas rate case from Johnson's campaign is here considered.
Superficial velocities are $\USG \approx \unitfrac[3.50]ms$ and $\USL \approx \unitfrac[0.25]ms$, and the test section is horizontal. 
These superficial velocities are acquired by a $\unitfrac[1.970]{kg}s$ liquid injection rate and a $\unitfrac[1.385]{kg}s$ gas injection rate at the inlet.
This is a low-amplitude wave case in which the two-fluid model is well behaved.
(If wave amplitudes grow bigger than they do here then eigenvalues turn locally complex at the wave tips due to high velocities.)
\\

\begin{figure}[ptb]%
\centering
\includegraphics[width=1\columnwidth]{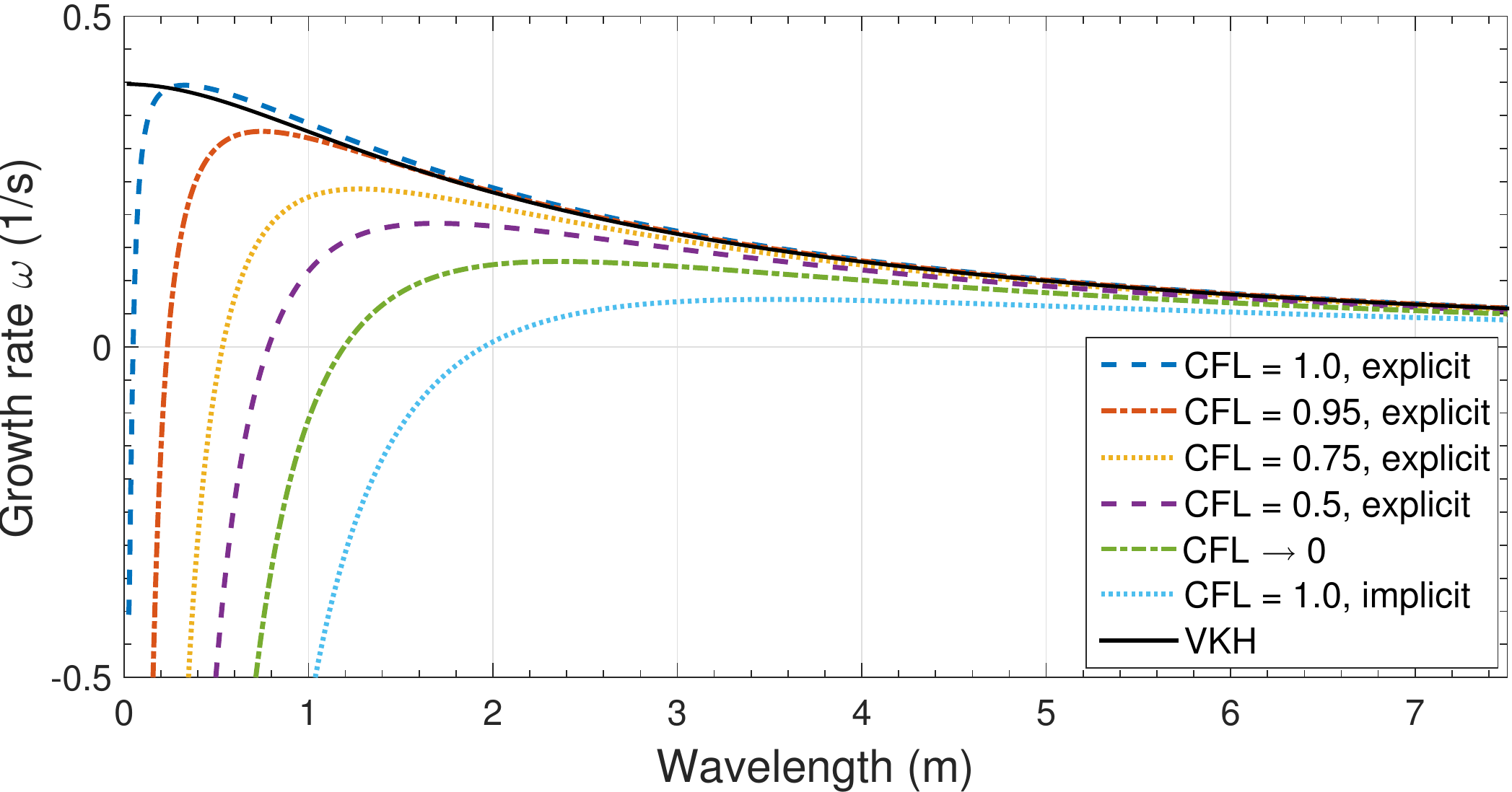}%
\caption{Linear wave growth of the two-fluid model \eqref{eq:base_model} and of Roe scheme representations of said model, computed form the equations in \autoref{sec:vonNeumann}. 
Flow state is as per the case description of \autoref{sec:exp:roll_wave}, with $\NJ{}\!\times\! \Nj{}=2\,000$ grid cells.
}%
\label{fig:exp:roll_wave:disc_stab}%
\end{figure}

We start by considering the linear growth of the two-fluid model.
By linear growth we mean the growth of small perturbations before nonlinear effects become considerable.
\autoref{fig:exp:roll_wave:disc_stab} presents graphs computed from the 
linear stability equations briefly presented in \autoref{sec:vonNeumann}. 
Growth rate $\omega$  
expresses the exponential/geometric perturbation growth of the differential/discrete two-fluid model.
Such graphs are drawn at virtually no computational expense and provide great insight into the accuracy of our discrete representations.
The figure shows the linear growth of the differential two-fluid model \eqref{eq:base_model} as a solid black line, together with the linear growth of the Roe scheme using $\NJ{}\!\times\! \Nj{}=2\,000$ grid cells at varying \CFL{} numbers as stippled lines.
Excluding the shortest wavelengths, the Roe scheme representations at \CFL{} close to one fit the VKH growth of the differential model nearly precisely. As pointed out in \autoref{sec:vonNeumann}, both the differential VKH growth graph and the discrete graph for explicit time integration and $\CFL{}=1.0$ would horizontal lines at $\omega = 0$ if the flow was at precise neutral stability.

Linear growth in the differential two-fluid model is strongest in the shortest wavelengths, with no finite wavelength of dominant growth. 
As a consequence, the wavelength distribution that emerges from out of the initial linear growth range will not be predisposed to any particular finite wavelength attributed to the model itself. 
Discrete representations, on the other hand, will suffer from numerical dissipation around the higher wavenumbers and will therefore emerge with a dominant wavelength. 

Effects from interface turbulence, wave breaking and surface tension, not included in the present two-fluid model, will in reality restrict the dominant wavenumber. 
Fundamental model limitations are also present as a result of the model's inherent long-wavelength assumption where dynamic pressure gradients are ignored. More of these effects ought to be included for a proper model/experiment comparison, as done in \cite{Holmaas_roll_wave_model}. 
The foci of the present section is foremost wave growth and numerical efficiency, and not experiment comparisons.
\\

Next we simulate the test case to evaluate the nonlinear evolution.
A pointwise random disturbance within $\pm 5\%$ of the injection rates is imposed at the inlet at each time step. 
$\NJ{}=20$ principal grid cells are used with $\Nj{}=100$ subgrid cells per principal cell.
The high number of subgrid cells is again chosen not for the purpose of spatial resolution, which would be fine also with fewer subgrid cells, but to allow for a hydraulic \CFL{} number near unity without violating the sonic time step restrictions.

\begin{figure}[ptb]%
\centering
\begin{subfigure}{\columnwidth}%
\includegraphics[width=\columnwidth]{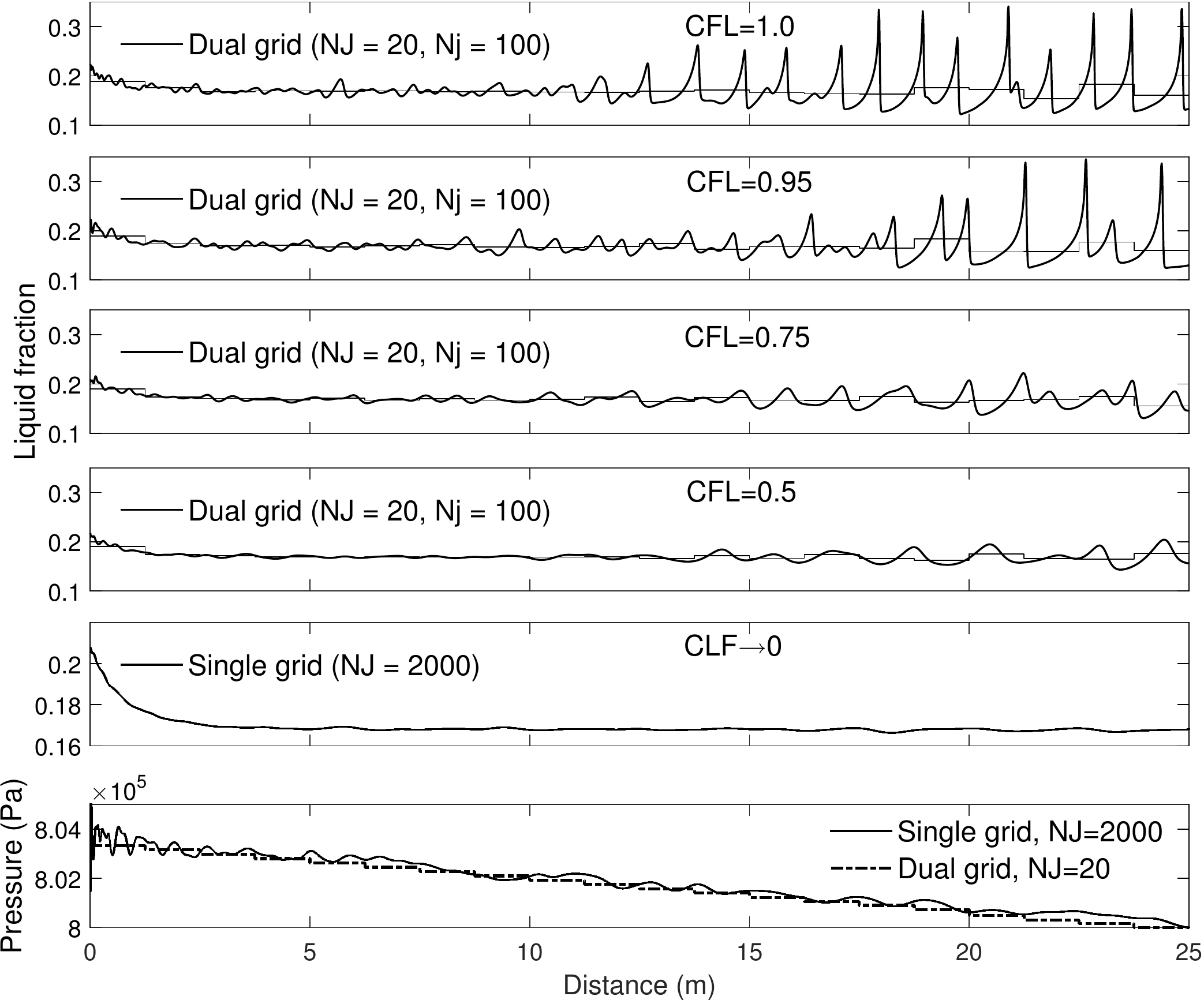}
\caption{Simulation snapshots.}
\label{fig:exp:roll_wave:snapshot}%
\end{subfigure}
\begin{subfigure}{.8\columnwidth}%
\includegraphics[width=\columnwidth]{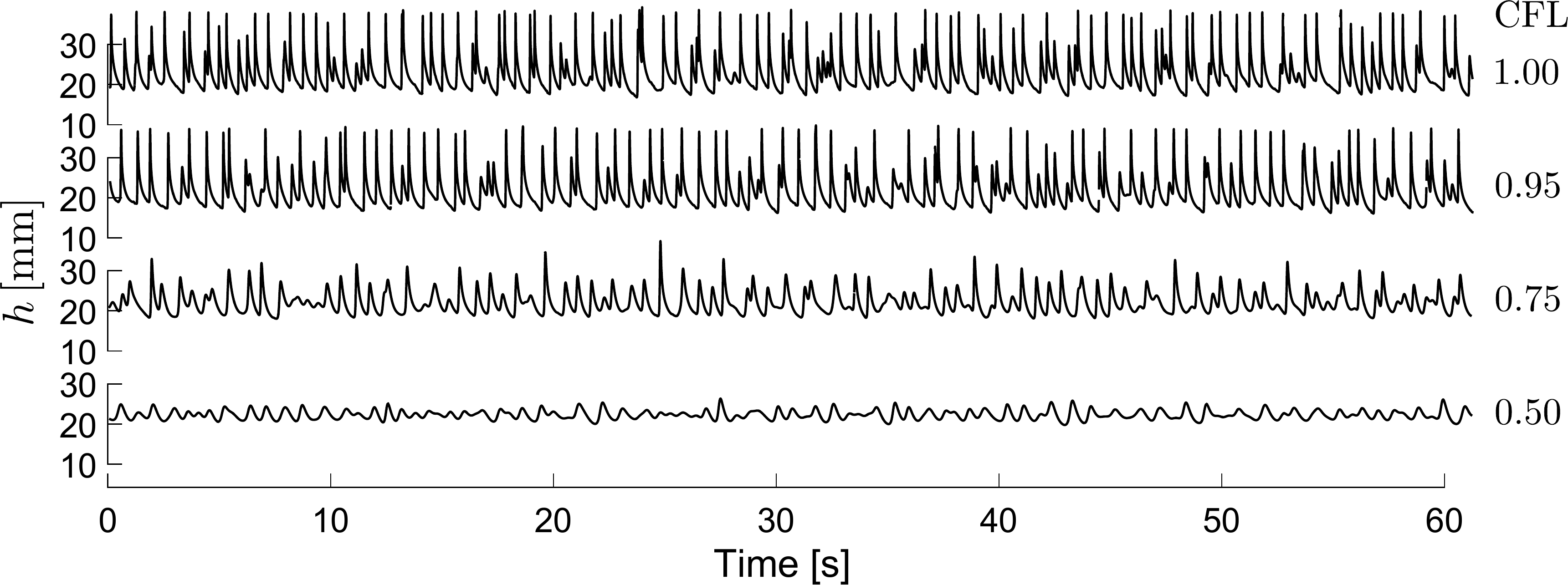}%
\caption{Time trace of level height $h$ sampled at $x=\unit[24]m$.}
\label{fig:exp:roll_wave:time_trace}%
\end{subfigure}
\begin{subfigure}{.9\columnwidth}%
\includegraphics[width=\columnwidth]{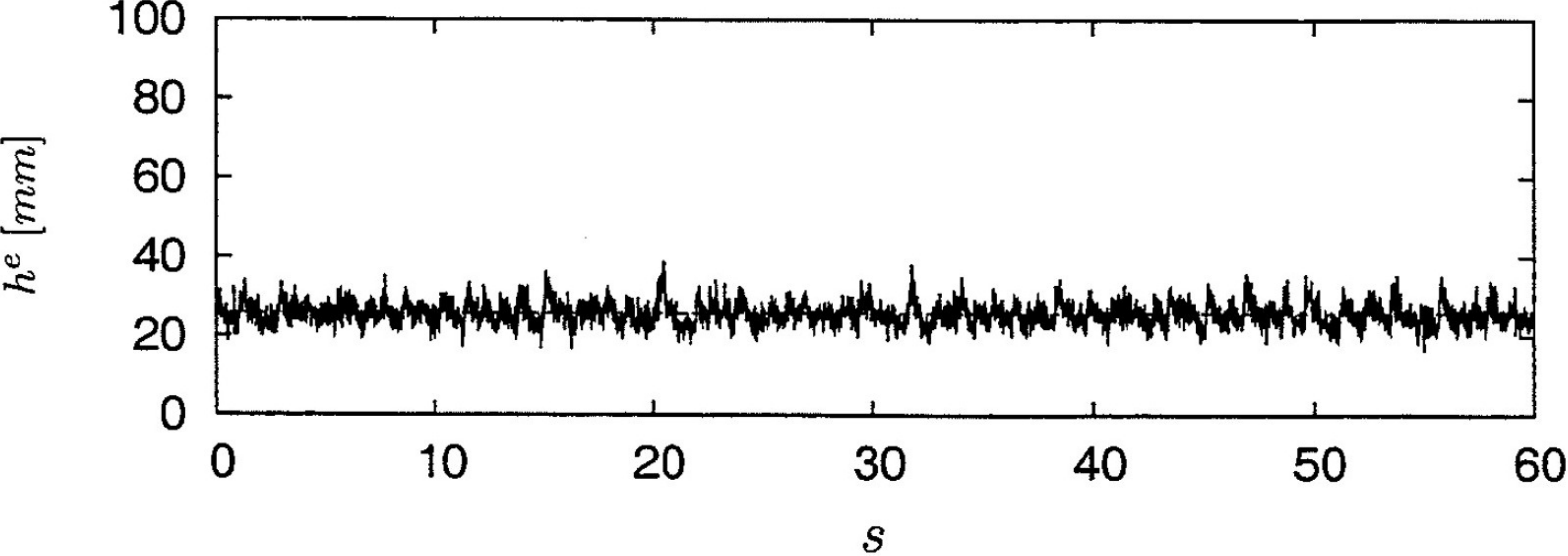}%
\caption{Experimental measurement; figure from \citep{George_PhD}.}
\label{fig:exp:roll_wave:time_trace_George}%
\end{subfigure}
\caption{
$\USG \approx \unitfrac[3.50]ms$, $\USL \approx \unitfrac[0.25]ms$, $\theta = 0\degree$.}
\label{fig:exp:roll_wave}%
\end{figure}

\autoref{fig:exp:roll_wave} shows simulations at four different \CFL{} numbers; $\CFL{} = 0.5$, $0.75$, $0.95$ and $1.0$.
High wavenumber growth is severely reduced for the lower \CFL{} numbers, as postulated in \autoref{fig:exp:roll_wave:disc_stab} and seen in the profile plots of \autoref{fig:exp:roll_wave:snapshot}.
Growth of all but the shortest wavelengths takes place in all simulations, but the rate of growth compared to the travelling time means that the time traces observed in \autoref{fig:exp:roll_wave:time_trace} differ in terms of amplitudes and frequencies. 
The wavelength distribution will remain inlet dependent if the growth and travelling time is insufficient for the waves to develop well into the nonlinear wave range.
For simple comparison,
a time trace of the relevant experiment from Johnson's campaign~\citep{George_PhD} is shown in \autoref{fig:exp:roll_wave:time_trace_George}. 
\\

Consider lastly the simulation efficiency.
Time steps for the $\CFL{} = 0.95$ simulation are around  $\dt = \unit[6\Em3]s$. 
The equivalent time step for a single grid  \HCU{} simulation with the same number of grid cells is $\dt = \unit[7.5\Em5]s$.
We might be tempted do say that the increase in computational efficiency is approximately $\unit[6\Em3]s/\unit[7.5\Em{5}]s = 80$, yet this does not take into account the accuracy-time step dependency shown earlier.  
Indeed, considering the tiny time step imposed by the sonic restrictions, linear growth with $2\,000$ \HCU{} cells on a single grid will be comparable to the graph marked `$\CFL{}\rightarrow0$' in \autoref{fig:exp:roll_wave:disc_stab}.
A snapshot of this simulation is shown in the second bottom panel of \autoref{fig:exp:roll_wave:snapshot}.
(The dual grid scheme gives an equally poor result if made to run at very small \CFL{} numbers.)
For the wave growth to be similar to the `$\CFL{}=0.75$' graph in \autoref{fig:exp:roll_wave:disc_stab}, and thus be comparable to the $\CFL{}=0.75$ plots of \autoref{fig:exp:roll_wave}, something like $\NJ{}=10\,000$ grid cells are needed.
The time step will then be around $\dt = \unit[1.5\Em5]s$, needing approximately $6.7\Ep8$ computations per simulated second.
This constitutes $2\,000$ times the computational expense of the equivalent dual grid simulation.

Backwards Euler (`implicit') time integration allows for time steps which are unrestricted by the sonic speeds. 
However, linear growth accuracy will then be comparable to the graph marked `CFL = 1.0, implicit' in \autoref{fig:exp:roll_wave:disc_stab}.
Higher order time integration schemes are likely to be less sensitive to the \CFL{} number.
Weakly implicit schemes~\citep{flaatten_WIMF}, which alleviates the sonic \CFL{} restriction while retaining explicitness in the hydraulic terms, is perhaps a better alternative.
\textcolor{\colorchanges}{Large time-step schemes 
\citep{Lindqvist_large_timestep_scheme,LeVeque_LTS_shocks_scalar}
can also be used to bypass stringent acoustic CFL-restrictions.}
\\

\section{Summary}
\label{sec:conclusions}
A dual grid arrangement has been proposed
which combines the \HCU{} scheme due to \cite{Evje_HCU_schemes} with the incompressible two-fluid model. 
The resulting method retains the robustness of the original \HCU{} scheme with regard to the acoustic evolution.
In addition, it allows for significant improvements in the accuracy and resolution of hydraulic waves at only moderate computational expense.
The increased efficiency is achieved chiefly by allowing 
the numerical time scales of sonic and hydraulic information to match.

Principal and subgrid models are in the proposed scheme coupled through the terms in the governing transport equations in a manner that does not generate numerical grid projection errors.
The incompressible subgrid methodology can be regarded by way of a scheme extension that does not necessitate significant alterations to a single grid scheme.
\\

Four different test cases have been presented. 
Applying an incompressible subgrid model to the shock tube and water faucet problems of \autoref{sec:exp:dispersive_tests} captured hydraulic shock details otherwise lost to diffusion. 
Pressure wave predictions also improved indirectly.

A surge wave--pressure wave case was presented in \autoref{sec:exp:surge_wave}. 
Here, the pressure-surge interactions were maintained when reducing the sonic resolution, improving computational efficiency by 1-2 orders of magnitude. 
Accurate hydraulic development and the essentials of the pressure-surge interactions were retained even 
if the principal grid was too coarse to resolve any acoustic wave front.
Computational efficiency was then increased by 2-3 orders of magnitude. 
Finally, an increase in computational efficiency by more than 3 orders of magnitude was demonstrated in the example roll-wave case of \autoref{sec:exp:roll_wave}.
Steady roll-wave flow is one of many situations where pressure dynamics play an idle role, acting as a steady background state.
We are usually not interested in the sonic propagation in such cases, but would still like the simulator to support compressible behaviour.

\section*{Acknowledgements}
This work is financed by The Norwegian University of
Science and Technology (NTNU) as a contribution the Multiphase
Flow Assurance programme (FACE.)
The author extends a personal thanks to Kontorbamse.


\section*{References}
\bibliographystyle{elsarticle-harv}
\bibliography{./refs_PhD}

\end{document}

%% file: usepackages.tex
\usepackage[utf8]{inputenc}
\usepackage[T1]{fontenc}

\usepackage{hyperref}
\hypersetup{colorlinks=true,citecolor=blue,linkcolor=blue,urlcolor=blue}

\usepackage[english]{babel}
\addto\extrasenglish{
  }

\usepackage{verbatim}
\usepackage{amssymb} 
\usepackage{amsthm} 
\usepackage{float}
\usepackage{enumitem}
\usepackage{graphicx} 
\usepackage{amsmath} 
\usepackage{nomencl}
\usepackage{gensymb} 

\usepackage{relsize}
\usepackage{units} 
\usepackage{bm} 
\usepackage{setspace}
\usepackage[font=normal]{caption}
\usepackage[belowskip=5pt,aboveskip=1pt,subrefformat=parens]{subcaption} 

\usepackage{tikz}
\usetikzlibrary{calc,fit,shapes.geometric,arrows,backgrounds,positioning}

\tikzset{
    >=stealth',
    punkt/.style={
           rectangle,
           rounded corners,
           draw=black, very thick,
           text width=6.5em,
           minimum height=2em,
           text centered},
    pil/.style={
           ->,
           thick,
           shorten <=2pt,
           shorten >=2pt,},
		remember picture			
}

\usepackage[mathlines]{lineno} 

\numberwithin{equation}{section} 
\numberwithin{figure}{section} 

\usepackage[final]{changes}
\usepackage{cancel}

%% file: newcommands.tex
\newcommand{\pp}{\partial}
\newcommand{\dd}{\mr d}
\newcommand{\pdiff}[2]{\frac{\partial #1}{\pp #2}}

\newcommand{\br}[1]{{ \left(  #1 \right) }}
\newcommand{\sqbrac}[1]{\left[ #1 \right]}

\newcommand{\lr}[3]{\left#1 #2 \right#3}
\newcommand{\an}[1]{\lr \langle {#1} \rangle}

\newcommand{\ie}{i.e.}
\newcommand{\egg}{e.g.}

\let\underscore\_
\renewcommand{\_}[1]{_\mathrm{#1} }

\let\texthat\^
\renewcommand{\^}[1]{^\mathrm{#1} }

\newcommand{\of}[1]{\br{#1}}

\newcommand{\ol}{\overline}
\newcommand{\mr}{\mathrm}
\newcommand{\mc}{\mathcal}
\renewcommand{\l}{\_{\ell}}
\newcommand{\g}{\_g}
\renewcommand{\k}{_\phaseindex}
\newcommand{\dx}{{\Delta x}}
\newcommand{\dX}{{\Delta X}}
\newcommand{\dt}{{\Delta t}}

\newcommand{\jm}{_{j-1}}
\newcommand{\jp}{_{j+1}}
\newcommand{\jmh}{_{j-\frac12}}
\newcommand{\jph}{_{j+\frac12}}

\newcommand{\kj}{_{\phaseindex,j}}
\newcommand{\kjp}{_{\phaseindex,j+1}}

\newcommand{\ljm}{_{\ell,j-1}}
\newcommand{\ljp}{_{\ell,j+1}}

\newcommand{\kjph}{_{\phaseindex,j+\h}}
\newcommand{\ljph}{_{\ell,j+\h}}

\newcommand{\lj}{_{\ell,j}}

\newcommand{\Jph}{_{J+\h}}
\newcommand{\Jmh}{_{J-\h}}
\newcommand{\J}{_J}
\newcommand{\Jp}{_{J+1}}

\newcommand{\kJph}{_{\phaseindex,J+\h}}
\newcommand{\lJph}{_{\ell,J+\h}}
\newcommand{\gJph}{_{\mr g,J+\h}}
\newcommand{\kJmh}{_{\phaseindex,J-\h}}
\newcommand{\kJ}{_{\phaseindex,J}}
\newcommand{\kJp}{_{\phaseindex,J+1}}

\newcommand{\lJ}{_{\ell, J}}
\newcommand{\gJ}{_{\mr g,J}}
\newcommand{\lJp}{_{\ell, J+1}}
\newcommand{\gJp}{_{\mr g,J+1}}
\newcommand{\h}{\frac12}

\newcommand{\n}{^n}
\newcommand{\nn}{^{n+1}}

\newcommand{\diff}[1]{\sqbrac{#1}}
\newcommand{\diffk}[1]{\diff{#1}\g\^\ell}
\renewcommand{\a}{a}
\newcommand{\al}{\a\l}
\newcommand{\ag}{\a\g}
\newcommand{\ak}{\a\k}
\newcommand{\area}{\mc A}
\newcommand{\Em}[1]{\times\! 10^{\text{-}#1}}
\newcommand{\Ep}[1]{\times\! 10^{#1}}

\newcommand{\alj}{a_{\ell,j}}
\newcommand{\aljp}{a_{\ell,j+1}}

\newcommand{\A}{A}
\newcommand{\Al}{\A\l}
\newcommand{\Ag}{\A\g}

\newcommand{\bv}{\bm v}
\newcommand{\bs}{\bm s}

\renewcommand{\bf}{\bm f}

\newcommand{\Qm}{\mc Q}
\newcommand{\ql}{\al u\l}
\newcommand{\qg}{\ag u\g}

\newcommand{\my}{g_y\diffk{\rho}}
\newcommand{\mx}{g_x\diffk{\rho}}
\newcommand{\HofAl}{\mc H}
\newcommand{\dHdAl}{\HofAl'}
\newcommand{\jac}{\pdiff \bf \bv}
\newcommand{\I}{\mathbb I}

\newcommand{\HCU}{\text{HCU}}
\newcommand{\CFL}{\text{CFL}}
\newcommand{\USL}{\ql/\area}
\newcommand{\USG}{\qg/\area}

\newcommand{\phaseindex}{k}
\newcommand{\symkappa}{\varkappa}
\newcommand{\imunit}{\mr i}
\newcommand{\eEuler}{\mr e}

\newcommand{\drhokdp}{\pdiff{\rho\k}p}
\newcommand{\drholdp}{\pdiff{\rho\l}p}
\newcommand{\drhogdp}{\pdiff{\rho\g}p}

\newcommand{\drhokdpEOS}{\pdiff{\rho\k}p\Big|_0}
\newcommand{\drholdpEOS}{\pdiff{\rho\l}p\Big|_0}
\newcommand{\drhogdpEOS}{\pdiff{\rho\g}p\Big|_0}

\newcommand{\Fkappa}{\kappa}
\newcommand{\F}{I}
\newcommand{\Roe}{\hat {\mathbb{A}}}
\newcommand{\Fp}{F_P}
\newcommand{\Fal}{F_{\Al}}
\newcommand{\CC}{\^C}
\newcommand{\UU}{\^U}
\newcommand{\Nj}{{N\!j}}
\newcommand{\NJ}{{N\!J}}

\newcommand{\labelref}[2][]{\hyperref[#2]{#1~\ref*{#2}}}

\newcommand{\colorchanges}{black}

\newcommand{\extension}{pdf}

\theoremstyle{plain}

%% file: abstract.tex
The speed of sound in two-phase pipe flow systems is often several orders of magnitude greater than the travelling speed of hydraulic information (volume fractions.)
Dynamically simulating such flows requires resolution of acoustic and hydraulic waves existing at vastly different spatial and temporal scales.
If simulated on the same numerical grid, the need for accuracy in hydraulic waves will necessitate an exaggerated resolution of acoustic waves. Likewise, time steps restricted by the speed of sound are small compared to the time scales active in hydraulic waves. 
This constitutes a waste of computational potential.
The method proposed herein decouples the hydraulic and acoustic scales, greatly improving computational efficiency.

The proposed dual grid method solves a four-equation compressible two-fluid model on a principal grid which robustly accounts for the pressure evolution and conserves mass and momentum. 
An incompressible two-equation model is at the same time solved on a finer grid, resolving the details of the hydraulic evolution.
Information from both model formulations is coupled through the terms of the governing transport equations, providing consistency between the grids.
Accurate and finely resolved schemes can then be employed for the incompressible two-fluid model without suffering from the time and stability restrictions otherwise enforced by acoustic waves. 
At the same time, the Hybrid Central-Upwind flux splitting scheme of \cite{Evje_HCU_schemes} allows for an explicit and
numerically robust treatment of the acoustic waves without losing hydraulic accuracy.

The dual grid method is tested against four dissimilar problems: A shock tube problem, the water faucet problem, a surge wave and pressure wave problem and a roll-wave case.
In all problems, the proposed scheme provided significant increases in computational efficiency and accuracy as compared with a single grid arrangement.